\documentclass[fleqn,usenatbib]{aa}
\usepackage[utf8]{inputenc}
\usepackage[varg]{txfonts}
\usepackage{color}
\usepackage{xcolor}
\usepackage[english]{babel}
\usepackage{fontenc,amssymb}
\usepackage{graphicx}
\usepackage[normalem]{ulem}
\usepackage[breaklinks=true]{hyperref}
\usepackage{ulem}
\usepackage[T1]{fontenc}
\usepackage[breaklinks=true]{hyperref}

\graphicspath{{Figures/}}

\newcommand{\FeII}{\hbox{{\rm Fe}{\sc \,ii}}}

\newcommand{\OII}{\hbox{{\rm O}{\sc \,ii}}}

\newcommand{\MgII}{\hbox{{\rm Mg}{\sc \,ii}}}

\newcommand{\mpy}{\hbox{M$_{\odot}$~yr$^{-1}$}}
\newcommand{\sigmpy}{\hbox{M$_{\odot}$~yr$^{-1}$~kpc$^{-2}$}}

\newcommand{\Vmax}{\hbox{$V_{\rm max}$}}
\newcommand{\Vout}{\hbox{$V_{\rm out}$}}
\newcommand{\thetam}{\hbox{$\theta_{\rm max}$}}

\newcommand{\kms}{\hbox{km~s$^{-1}$}}

{}
{}
{}
{}
{}
{}
{}

\begin{document}
\title{MusE GAs FLOw and Wind (MEGAFLOW) XI. Scaling relations between outflows and host galaxy properties~\thanks{
Based on observations made at the ESO telescopes at La Silla Paranal Observatory under programme IDs 094.A-0211(B), 095.A-0365(A), 096.A-0164(A), 097.A-0138(A), 099.A-0059(A), 096.A-0609(A), 097.A-0144(A), 098.A-0310(A), 293.A-5038(A).}}
\author{Ilane~Schroetter\inst{1} \thanks{ilane.schroetter@gmail.com}\and
        Nicolas~F.~Bouch\'e$^{2}$ \and
        Johannes~Zabl$^{2,6}$\and
        Martin~Wendt$^{3}$\and
        Maxime~Cherrey$^{2}$\and
        Ivanna~Langan$^{4,2}$\and
        Joop~Schaye$^{5}$ \and
        Thierry~Contini$^{1}$
        }
\institute{Institut de Recherche en Astrophysique et Planétologie, Université de Toulouse, CNRS, CNES, UPS, 9 Av. du colonel Roche, 31028 Toulouse Cedex 04, France \and
{Centre de Recherche Astrophysique de Lyon (CRAL) UMR5574, Univ Lyon1, Ens de Lyon, CNRS, F-69230 Saint- Genis-Laval, France}\and
{Institut f\"ur Physik und Astronomie, Universit\"at Potsdam, Karl-Liebknecht-Str. 24/25, 14476 Golm, Germany}\and
{European Southern Observatory, Karl-Schwarzschild-Str. 2, D-85748, Garching, Germany}\and
{Leiden Observatory, Leiden University, PO Box 9513, 2300 RA Leiden, The Netherlands}\and
{Institute for Computational Astrophysics and Department of Astronomy \& Physics, Saint Mary’s University, 923 Robie Street, Halifax, Nova Scotia, B3H 3C3, Canada} 
}

\label{firstpage}
\abstract
{Absorption line spectroscopy using background quasars can provide strong constraints on galactic outflows. In this paper, we investigate possible scaling relations between outflow properties, namely outflow velocity \Vout, the mass ejection rate $\dot M_{\rm out}$, and the mass loading factor $\eta$ and the host galaxy properties, such as star formation rate (SFR), SFR surface density, redshift, and stellar mass  using galactic outflows probed by background quasars from MEGAFLOW and other surveys.

We find that $V_{\rm out}$ ($\eta$) is (anti-)correlated with SFR and SFR surface density.
We extend the formalism of momentum-driven outflows of Heckman et al. to show that it applies not only to down the barrel studies but also to winds probed by background quasars, suggesting a possible universal wind formalism.

Under this formalism, we find a clear distinction between ``strong'' and ``weak'' outflows
where 
``strong'' outflows seem to have tighter correlations with galaxy properties (SFR or galaxy stellar mass) than ``weak'' outflows.}
\titlerunning{MEGAFLOW XI. Scaling relations between outflows and host galaxy properties}
\maketitle
\section{Introduction}
In the era of modern cosmology with well determined cosmological parameters, the process (or processes) responsible for the low efficiency of galaxy formation is still unknown.  Indeed,
the efficiency of galaxy formation, defined as the fraction of baryons {in galaxies} relative to the amount of baryons available for a given cosmology, is low, ranging from a few percent to  20 percent
\citep[e.g.][]{BehrooziP_13a,moster_13}.
The peak occurs at around the Milky Way DM mass (or $10^{12}$ M$_\odot$).
At low masses, it is common to invoke feedback processes such as supernovae explosions \citep[e.g.][]{DekelA_86a},
radiation pressure \citep[e.g.][]{MurrayN_05a, HopkinsP_12a}, cosmic rays or stellar winds to account for the low efficiency of galaxy formation.
At higher masses, feedback processes from active galactic nuclei (AGN) are thought to play a major role \citep[e.g.][]{SilkJ_Rees_98}. Both SN-driven and AGN-driven outflows would drive baryons out of galaxies back into the circum-galactic medium (CGM).

The CGM, loosely defined as the region (within the virial radius, or $<$100-150 kpc) surrounding galaxies, is where the
signatures of these complex feedback processes for galaxy evolution are to be found, including gas accretion from the intergalactic medium (IGM). 
Thus, observations of the CGM are crucial in order to put constraints on galaxy formation numerical models. 
However, the CGM is difficult to observe directly because the gas density is orders of magnitude lower than the interstellar medium of the host galaxy. Fortunately,
bright background sources like quasars are effective  probes to study the CGM because they
(i) are sensitive to low gas (or column) densities around foreground objects and (ii) unveil the presence as well as kinematics of gaseous halos around any type of galaxy, irrespective of their luminosity or star-formation rate (SFR) 
\citep[e.g.][]{BoucheN_07a,BoucheN_12a,TurnerM_14,KacprzakG_14a, SchroetterI_15a,SchroetterI_16a, muzahid_15, RahmaniH_18, MaryD_20}. 

Compared to traditional techniques requiring imaging and expensive spectroscopic campaigns \citep[e.g.][]{BergeronJ_86a, SteidelC_95a, NielsenN_13a}, 
Integral Field Units (IFU) 
combined with background quasars
provide the most efficient way
to  study the properties of gaseous halos surrounding  galaxies because they yield simultaneously the redshifts of all galaxies in the field-of-view, thereby allowing for a rapid identification of absorption-galaxy pairs \citep[e.g.][]{BoucheN_12a, SchroetterI_15a, SchroetterI_16a, SchroetterI_19a, ZablJ_19a, MartinC_19, MuzahidS_20}. 
Wide-field IFUs like MUSE \citep{BaconR_06a, BaconR_10a,BaconR_15a}
are especially important given that they can study absorption-galaxy pairs up to 200--300 kpc, going beyond a typical virial radius at intermediate redshifts.

In the past years, several IFU surveys have yielded large samples of absorption-galaxy pairs such as MUSEQuBES \citep{MuzahidS_20}, CUBS \citep{ChenH_20}, MAGG \citep{LofthouseE_20}. In particular,
our MUSE GAs FLow and Wind survey (MEGAFLOW) has yielded a sample of {more than} 100 \MgII\ absorber-galaxy pairs at $0.4<z<1.5$ (Bouché et al., in prep.).
A clear result from this survey and others \citep[e.g.][]{BordoloiR_11a,BoucheN_12a,KacprzakG_11b,SchroetterI_15a,HoS_17a, LanT_17, LundgrenB_21}  is that the cool CGM is anisotropic with an excess \MgII\ absorption along the minor and major axes of star-forming galaxies (SFGs), indicating two physical mechanisms responsible for the \MgII\ absorption around galaxies, the former being outflows \citep{SchroetterI_16a,SchroetterI_19a} and the latter being extended gaseous disks \citep{ZablJ_19a}. This dichotomy is supported by the absorption kinematics 
with respect to the host 
\citep{SchroetterI_19a,ZablJ_19a}, see also \citet{KacprzakG_11b, NielsenN_15a, BordoloiR_11a, MartinC_19, LundgrenB_21}, 
and allows one to study the properties of outflows (kinematics, mass outflow rates, etc.).

While numerous studies exist on galactic outflows using traditional spectroscopy (aka the down-the-barrel technique) \citep[e.g.][]{GenzelR_11a, MartinC_05a, HeckmanT_15a} 
 only a few groups have used the background QSO technique to put constraints on outflow properties like the outflow velocity such as 
 (i) The KBSS survey \citep{SteidelC_kbss_14}, a galaxy redshift survey around 15 luminous QSO; 
(ii) the COS-burst survey \citep{HeckmanT_17a} around 17 low redshift starburst galaxies 
;
(iii) the Keck survey for \MgII\ halos around 50 $z\sim0.2$ normal SFGs \citep{MartinC_19}. 

In this paper, we focus on the properties (outflow velocity, ejected mass rate and mass loading factor) of galactic outflows and investigate possible scaling relations with the properties of the host galaxy. 
The paper is organized as follows: 
in section \S~\ref{section:data}, we briefly present the MEGAFLOW sample. 
In section \S~\ref{section:samples_scaling_relations}, we investigate the different wind properties and compare them with the host galaxy properties, adding other studies in search of possible scaling relations.
In \S~\ref{section:samples_mechanisms} we discuss the possible origin of outflow mechanisms and {in} \S~5 is where {are} discussion and conclusions.

Throughout the paper we use a 737 cosmology ($H_0=70$~\kms~Mpc$^{-1}$, $\Omega_{\rm m}=0.3$, and $\Omega_{\Lambda} = 0.7$) and a \citet{ChabrierG_03a} stellar Initial Mass Function (IMF).

\section{Data}
\label{section:data}

\subsection{The MEGAFLOW survey}
 
The MusE GAs FLOw and Wind (MEGAFLOW) survey consists of 22 quasar fields selected to have {\it multiple} strong \ion{Mg}{II} absorption lines (rest-frame equivalent width $W_r^{\lambda2796} > 0.5 - 0.8$\AA) in the \citet{ZhuG_13a} catalog based on the Sloan Digital Sky Survey (SDSS; \citealt{ross_12}; \citealt{alam_15}),
which resulted in 79 strong \ion{Mg}{II}  absorbers\footnote{{79 absorbers in the data release (DR)1 of MEGAFLOW, with now up to 127 which constitutes DR2}}.
The survey was designed to study the properties of gas flows surrounding star-forming galaxies (SFGs)  detected using the Multi Unit Spectroscopic Explorer (MUSE, \citealt{BaconR_10a}) on the Very Large Telescope (VLT).  
For each quasar, we carried out high-resolution spectroscopic follow-up observations with the Ultraviolet and Visual Echelle Spectrograph (UVES, \citealt{DekkerH_00a}).

We refer the reader to \citet{SchroetterI_16a} (hereafter paper I) for a more detailed understanding of the observational strategy, to \citet{ZablJ_19a} (hereafter paper II) for data reduction. 

In \citet{SchroetterI_19a} (hereafter paper III), we constrained outflow properties of 27 $z\approx 1$ host galaxies, namely, we constrained the outflow velocity \Vout, the mass outflow rate $\dot M_{\rm out}$ and the mass loading factor
$\eta=\dot M_{\rm out}/\rm SFR$, i.e. the ratio between the mass ejected rate and the Star Formation Rate (SFR).
In this paper, we seek to address whether these
 outflow properties follow scaling relations {with} host galaxy properties (e.g. SFR, stellar mass ($M_*$), redshift).

\subsection{Previous studies on galactic winds}
\label{subsection:previous_studies}

In order to {augment our} results of paper~III with other {outflow} studies 
which also used
 the background quasar technique, {we will include}
 {{(i)
  \citet[][hereafter B12]{BoucheN_12a} who use a catalog of 11 galaxy-quasar pairs\footnote{of which 5 are in a configuration for wind studies} from a combination of SDSS, Keck LRIS and OTA observations
  (ii) the 4 wind cases of the SIMPLE sample \citep{SchroetterI_15a} which are a combination of VLT/SINFONI and UVES observations.}
(i{ii}) 
\citet[][hereafter H17]{HeckmanT_17a} who built the COS-burst catalog containing 17 starburst galaxies  selected from the  SDSS DR7  and QSOs from GALEX DR6 catalog;
(i{v})
\citet[][hereafter M19]{MartinC_19} who used a sample of 50\footnote{{of which 30 have \MgII\ velocities, of which only 16 have an azimuthal angle with their quasar suitable for wind studies}} galaxy-quasar pairs at $z\approx0.2$.

To compare wind properties from background quasars to the more common down-the-barrel technique, we will also use
({v}) 
\citet[][hereafter H15]{HeckmanT_15a} who used both the COS-FUSE and COS-LBA surveys \citep[][respectively]{grimes_09, AlexandroffR_15a}; and
(v{i}) \citet[][hereafter {P23}]{PerrottaS_23} who used a sample of {14} starbursts based on the SDSS DR8 catalog.
Table~\ref{table:other_wind_studies}
summarizes the characteristics of these surveys.
As there are many down-the-barrel studies, we choose only those {which reported} outflow properties like the ejected mass {outflow} rates and mass loading factor. 
The two down-the-barrel  studies H15 and P2{3} were chosen for the following reasons: 
(1) they focus on low-redshift galaxies ($z<0.8$) and are thus complementary to our $z\approx1$ sample;
(2) the stellar mass range probed is similar to ours ($\log M_*/\rm M_\odot \approx 7 - 12$); and (3) the galaxies are starbursting and are thus complementary to our more ``normal'' star-forming galaxies. 
Table~\ref{table:other_wind_studies} summarizes the general properties of {each} study. 
Concerning SFRs of the MEGAFLOW galaxies, paper III discussed the possible bias between using SED fitting and [\OII] emission. Comparing SFRs from different sample, they find that there is no systematic bias between both methods.

\begin{table*}
\centering
\caption{Summary of other wind studies.}
\label{table:other_wind_studies}
\begin{tabular}{lcccccc}
\hline
Paper & $N_{\rm gal}$ &  $z_{\rm gal}$    & SFRs & Galaxy type    & $\log(M_{\star}/\rm M_\odot)$ & Method  \\
(1)   &  (2)          & (3)               & (4)  & (5)        & (6)            & (7)  \\
\hline
Paper III & 27 &$0.5<z<1.5$       & [\OII]  & SFGs      & $8.5 - 11$ & Background QSO \\
{B12} & {5} &$z\approx0.1$       & {SED}  & {SFGs}      & $10 - 11$ & {Background QSO} \\
S15 & 4 &$0.5<z<1.5$       & [\OII]  & SFGs      & $8.5 - 11$ & Background QSO \\
H17 & 17 &$z<0.2$       & SED  & Starbursts     & $10 - 11$ & Background QSO \\
M19 & {16} &$z<0.4$       & SED  & SFGs      & $9 - 11$ & Background QSO \\
H15 & {32} &$z<0.2$       & SED  & Starbursts     & $7 - 11$ & Down-the-barrel \\
P23 & 14 &$z\approx0.5$       & SED  & Starbursts      & $10.5 - 11.2$ & Down-the-barrel\\
\hline
\end{tabular}\\
{
(1) Study name, Paper III for \citet[][]{SchroetterI_19a}, {B12 for \citet{BoucheN_12a}}, S15 for the SIMPLE wind cases in \citet{SchroetterI_15a}, H15 and H17 for \citet{HeckmanT_15a, HeckmanT_17a}, respectively, M19 for \citet[][]{MartinC_19}
and P23 for \citet[][]{PerrottaS_23};
(2) Sample size in galaxy number.
(3) Redshift range;
(4) Method to estimate galaxy SFR;
(5) Type of galaxies;
(6) Galaxies stellar mass range;
(7) Method used to constrain outflow properties.
}
\end{table*}

\subsection{Sample general properties}

{To better understand the evolutionary state of the various samples, we show the selected literature galaxy samples} and MEGAFLOW+SIMPLE \citep[][]{SchroetterI_15a} galaxies  relative to the main sequence between SFR and galaxy stellar mass $M_*$ in Fig.~\ref{fig:ms}. 
In order to {account for the redshift evolution},
we choose to use the relation derived by \citet[][]{BoogaardL_18} and ``normalized" for a redshift $z=0.2$.
In this figure, we clearly see that H15 galaxies are in the high SFR part, close to starburst galaxies, whereas MEGAFLOW and M19 
galaxies appear to be mainly located on the main sequence at $z=0.2$.
{In Fig.~\ref{fig:ms}, red symbols represent}
 either starbursts (for {P23} and H17) or ``strong'' outflow cases (for H15, M19 and Paper III galaxies) as described in \S~\ref{section:formalism}.

\begin{figure}
  \centering
  \includegraphics[width=8cm]{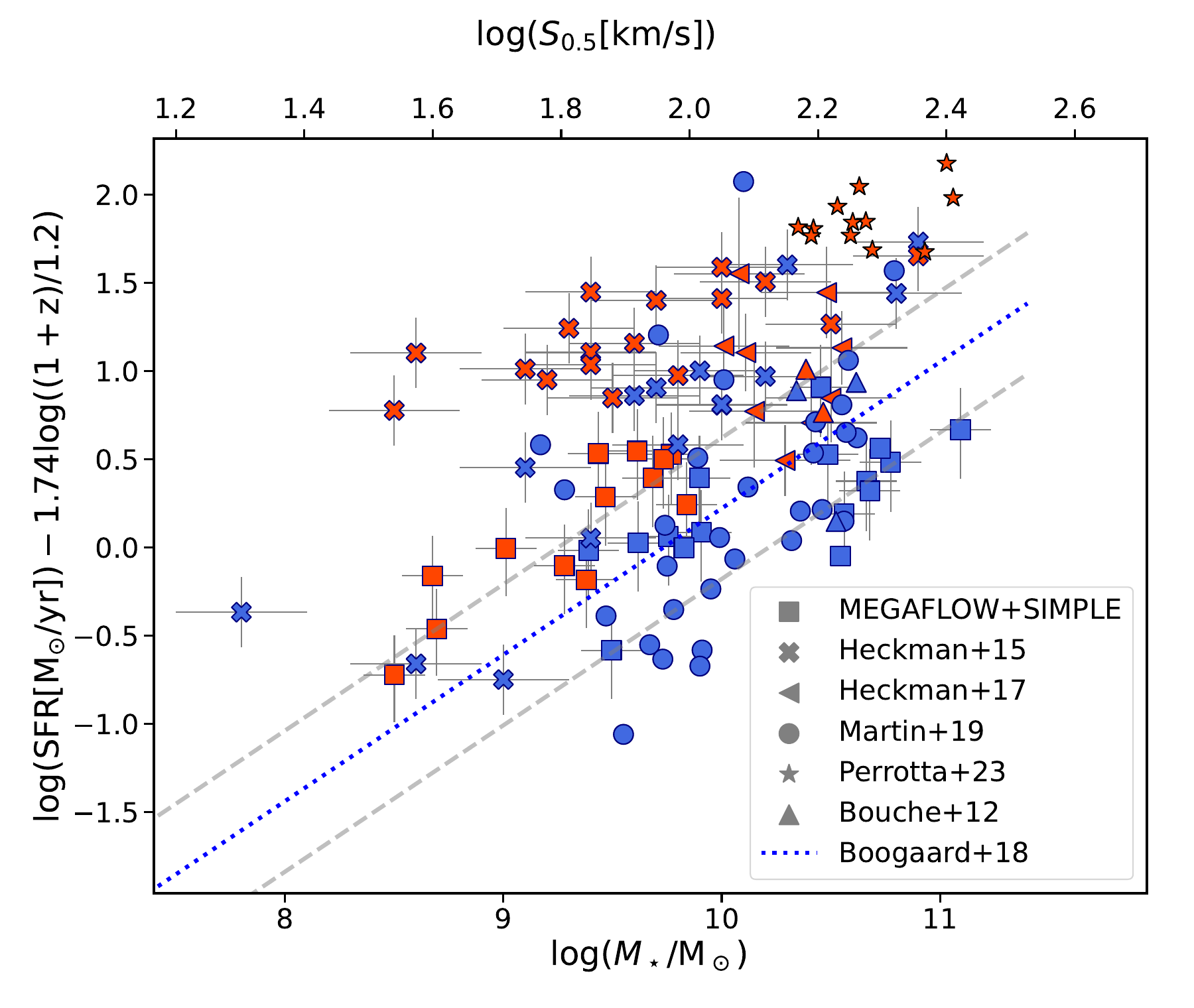}
  \caption{SFR$-M_\odot$ main sequence. Data points are normalized for redshift evolution to $z=0.2$.
  All the SFR are also corrected to have the same Initial Mass Function (a \citealt{ChabrierG_03a} IMF).
  The  \citet[][]{BoogaardL_18} MS relation is represented by the dotted blue line along with its intrinsic scatter
  of 0.4 dex (gray dashed lines).
  As described in the text (in \S~\ref{section:formalism}), red (blue) galaxies are considered as ``strong'' (``weak'') outflow cases.
    }
  \label{fig:ms}
\end{figure}

 \begin{figure*}[ht!]
  \centering
  \includegraphics[width=16.0cm]{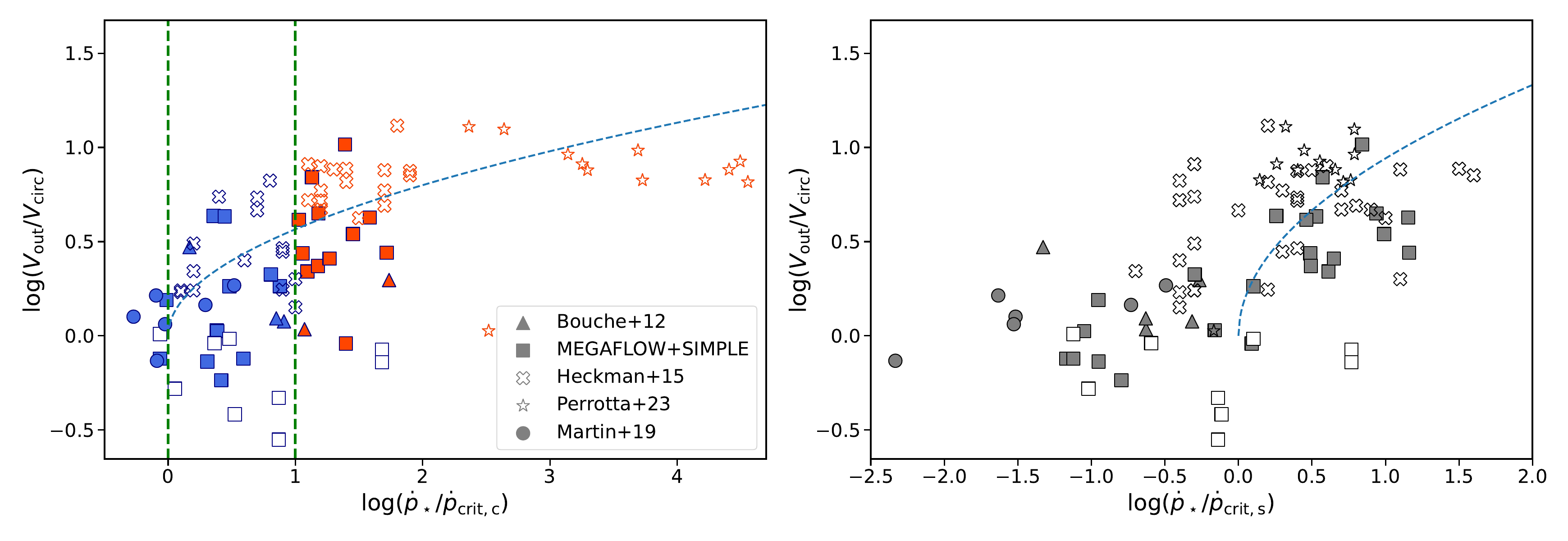}
  \caption{
  Normalized outflow velocity ($\rm V_{\rm out}/\rm V_{\rm circ})$ as a function of the ratio of the amount of momentum flux supplied by star formation to the value needed to overcome gravity and push the outflowing gas {for a cloud model (left) and for a shell model (right)}. 
  This figure is similar to Fig. 9 in H15 study.
  Squares, crosses{, triangles, circles and stars} represent MEGAFLOW, H15{, B12, M19 and P23} galaxies, respectively.
  {Empty symbols represent down-the-barrel data whereas filled ones are from background quasars studies.}
  {On the left panel, }the blue (red) color shows the ``weak'' (``strong'') outflows.
  {On the left (right) panel, }the dashed curve is derived from the equation of motion {for a cloud (shell)} model ({see text}).
  White squares correspond to less reliable MEGAFLOW outflow properties (more details in \citet[][]{SchroetterI_19a}).
  }
  \label{fig:momentum}
\end{figure*}
\section{Results}
In order to  {compare wind studies
 using background quasars to other types of studies,}
we first need to {establish} a 
{common} framework. 
In {particular}, H15 makes a {distinction} between ``strong'' and ``weak'' outflows based on the
wind momentum compared to the momentum injection rate from SFR {and extend their formalism to winds probed by background quasars sight-lines.}

\subsection{Wind formalism}
\label{section:formalism}

{Following H15,
in the case of} momentum-driven winds where the momentum injection rate $\dot p_\star$ is supplied by the star-forming or starburst galaxy, the outward force from the momentum injection is countered by gravity.
For an outflow to develop, the outward force ought to be greater than gravity, defining a critical momentum injection rate $\dot p_{\rm crit}$ (e.g. H15). 

{For a cloud outflow model} (to be consistent with
 \citealt{BoucheN_12a} and \citealt{SchroetterI_15a}),  the outward force is the pressure $P_{\rm w}$ times the cloud area $A_{\rm c}$, 
\begin{equation}
F_{\rm w}=P_{\rm w}\,A_{\rm c}=\frac{\dot p_\star}{\Omega_{\rm w} r^2} \,A_{\rm c},\label{eq:Fout}
\end{equation}
where $\Omega_{\rm w}$ is the wind solid angle and $r$ the location of the wind, 
while the inward gravitational force is
\begin{equation}
F_{\rm g}=\frac{G\,M(<r)}{r^2}m_{\rm c}=\frac{v^2_{\rm circ}}{r}m_{\rm c},\label{eq:Fgrav}
\end{equation}
where $G$ is the gravitational constant and $v_{\rm circ}^2$ the galaxy circular velocity.
 For a cloud of mass $m_{\rm c}$ and area $A_{\rm c}$,  the critical momentum flux $\dot p_{\rm crit, c}$
 is given by $F_{\rm w}\geq F_{\rm g}$, or $\dot p_{\rm crit, c}(r) A_{\rm c}\geq \Omega_{\rm w} v_{\rm circ}^2\, r\, m_{\rm c}$, such that, if one writes the cloud mass as 
$m_{\rm c}=A_{\rm c}(r)\,N_{\rm c}(r)\,\langle m \rangle$ where $N_{\rm c}(r)$ is the cloud column density and $\langle m \rangle$ {is }the mean mass per particle,
the critical momentum flux is
 \begin{eqnarray}
         \dot p_{\rm crit, c}(r)&=&
         \Omega_{\rm w}\,v_{\rm circ}^2\,r\,m_{\rm c}/A_{\rm c}\nonumber\\
         &=&
         \Omega_{\rm w}\,v_{\rm circ}^2\,r\,N_{\rm c}(r)\,<m>,\label{eq:pcrit}
 \end{eqnarray}

This critical momentum flux  required in order to have a net outward force on an outflowing cloud is [H15]:
\begin{equation}
\dot p_{\rm crit, c}(r)=10^{33.9}{\rm dyn}\,\frac{\Omega_{\rm w}}{4 \pi}\,\frac{N_{\rm c}(r)}{10^{21}{\rm cm}^{-2}}\,\frac{r}{1{\rm kpc}}\left(\frac{v_{\rm circ}}{100~\kms}\right)^2.
\end{equation}
In {other} words, winds will only develop when $\dot p_\star>\dot p_{\rm crit, c}$.

Comparing this critical momentum flux $\dot p_{\rm crit, c}$ to the momentum injection rate $\dot p_\star$, which is $\dot p_\star = 4.8\times10^{33}~{\rm SFR}$~dynes, one can distinguish between ``weak'' and ``strong'' outflows.  H15 defined ``weak'' winds when $0<\log(\dot p_{\rm *}/\dot p_{\rm crit, c})<1.0$ and ``strong'' winds when $\log(\dot p_{\rm *}/\dot p_{\rm crit, c})>1.0$.  
{This means that these regimes are the two cases where their momentum flux is larger or lower than ten times the critical momentum flux required to have a net outward force on an outflowing cloud.}
{H15 made this arbitrary limit to where the ``strong'' outflow seems to carry a significant amount (of the order of unity) of the momentum flux available from the starburst.}

From Eqs~\ref{eq:Fout}--\ref{eq:Fgrav},   the equation of motion for such a cloud launched from $R_0$ is
\citep[e.g.][]{MurrayN_05a,HeckmanT_15a}:
\begin{eqnarray}
 \frac{1}{2}v^2
&=&\int_{R_0}^{r} {\rm d}r \left ( \frac{\dot p_\star}{\Omega_{\rm w}r^2}\frac{A_{\rm c}}{m_{\rm c}}-\frac{v_{\rm circ}^2}{r} \right ) = v_{\rm circ}^2 \int_{R_0}^{r} {\rm d}r\left(\frac{R_{\rm g}}{r}-1\right) \frac{1}{r} \\
v(r)&=& \sqrt{2}\, v_{\rm circ}\sqrt{\left(1-\frac{R_0}{r}\right)\frac{R_{\rm g}}{R_0}-\ln\left(\frac{r}{R_0}\right)} \label{eq:motion}
\end{eqnarray}
where $R_{\rm g}={\dot p_{\star}}/({\Omega_{\rm w}\,v_{\rm circ}^2 \,N_{\rm c}<m>}) $ defines the radius at which the velocity peaks at $v_{\rm max}$
\begin{eqnarray}
v_{\rm max} 
&=&\sqrt{2}v_{\rm circ}\sqrt{\left(\frac{\dot p_\star}{\dot p_{\rm crit, c}}\right)-1-\ln\left(\frac{\dot p_\star}{\dot p_{\rm crit, c}}\right)}\\
&\simeq& \sqrt{2}v_{\rm circ}\sqrt{\frac{R_g}{R_0}}\qquad\hbox{can be $\gg \sqrt{2}\,v_{\rm circ}$}\nonumber
\end{eqnarray}
where  $R_{\rm g}/R_0=\dot p_{\star}/\dot p_{\rm crit, c}(R_0)$ \citep[defined as $R_{\rm crit}$ in][]{HeckmanT_15a}.

Note that this formalism {makes a number of implicit assumptions. First,} it assumes that the potential is that of an isothermal sphere, $v_{\rm circ}^2=\frac{G\,M(<r)}{r}=2\sigma^2$ , which implies that $v_{\rm circ}$ is independent of radius. {Second,} the expression for the critical momentum
injection rate $\dot{p}_{\rm crit}$  is estimated at $r=R_0$ where $R_0$ is the launch radius. H15 uses $R_0=r_\star$=1~kpc.

{In the case of background sight lines, the critical momentum $\dot p_{\rm crit, c}$ is only evaluated at $r=b$, where $b$ is the impact parameter. However, }assuming that outflows  are mass conserving, i.e. $\rho(r) r^2$ is constant, such that $N(r) r$ is conserved, then Eq.\ref{eq:pcrit} implies that $\dot p_{\rm crit, c}$ is independent of radius, $\dot p_{\rm crit, c}(b)=\dot p_{\rm crit, c}(R_0)$, provided that the circular velocity $v_{\rm circ}$  is roughly constant.

{Following H15, we also looked at a shell model for the outflowing gas\footnote{{The shell wind model  assumes that the outward force $p_\star$ is larger than the shell gravity $M_{\rm s}\,v_{\rm circ}^2/r_{\rm s}$ for a shell mass $M_{\rm s}$ at radius $r_{\rm s}$.}} using the critical momentum injection $\dot p_{\rm crit, s} = f_{\rm s}v_{\rm cir}^4/G$, where
the shell mass fraction $f_{\rm s}\equiv M_{\rm s}/M(<r)$ is $\sim0.1$. 
Fig.~\ref{fig:momentum}(right)
shows that a shell model (Eq.~6 in H15) does not match the data compared to the cloud model shown in the left panel. We thus stick with the cloud model as it appears to better describe the data. } 

\subsection{A universal formalism?}

{In order to test this formalism, we first} compare  the outflow velocity to the wind strengths
in Figure~\ref{fig:momentum}.
This figure shows
the normalize outflow velocity $v_{\rm out}/v_{\rm circ}$ as a function of $\log(\dot p_{\rm *}/\dot p_{\rm crit, c})$ ratio for the MEGAFLOW (paper III), H15, M19 \footnote{Note that for M19 systems, we only use those with a \MgII\ REW $W_r^{\lambda 2796}$ larger than 0.5~\AA\ in order to match the MEGAFLOW sample selection and have a consistent estimation of the hydrogen column density, which reduces the number of galaxies from 16 to 8 for this sample.} and P23 galaxies for a cloud and a shell model on left and right panels, respectively.
{For the galaxies in paper~III, we use $v_{\rm circ}\approx S_{0.5}=\sqrt{0.5\Vmax^2+\sigma^2}$ where \Vmax\ is the ``intrinsic'' galaxy rotational velocity (corrected for the galaxy inclination), while H15 used the observed rotational velocity.
Finally, we use the bi-conical shape of our outflows with a cone opening angle of $2\theta=60^\circ$ because of the {azimuthal bi-modality in \MgII{}
\citep{BordoloiR_11a,BoucheN_12a,SchroetterI_15a,LundgrenB_21}.}
This leads to a {downward} correction of H15 estimations of their ejected mass rate~\footnote{ 
To get from a spherical to a bi-conical outflow with opening angles of 60 degrees geometry, the reduction is approximately 8.}.
We also use their updated outflow velocities values \citep[][]{HeckmanT_16a}.
{In addition, we investigated for a possible correlation between sight-line impact parameter and whether the outflow is classified as ``strong'' or ``weak'' but found none.}

In Fig.~\ref{fig:momentum} and subsequently, we  use  blue (red) symbols for ``weak'' (``strong'') outflows when showing MEGAFLOW, H15, M19 and P23 results. 
Also, throughout the paper, white squares correspond to galaxies} in Paper III where the wind model does not fit the {spectra} convincingly\footnote{The reasons for each case are described in paper III, one of the main reason is that the absorption system has multiple component and thus is too complex to be reproduced by the simple wind model as the absorption most likely is a combination of multiple galaxy contribution.}. 
{In addition to differentiating ``weak'' and ``strong'' outflows in blue and red, respectively, we also use empty markers for ``down-the-barrel'' cases, namely H15 and P23, throughout the paper for a clear distinction between both methods.}

\subsection{Wind scaling relations}
\label{section:samples_scaling_relations}
Because galaxy properties like SFR and mass { are thought} to be directly linked with properties of galactic winds \citep[e.g.][]{HeckmanT_00a, MartinC_05a, HopkinsP_12a, NewmanS_12a, HeckmanT_15a}
, we will investigate the relations, if any, between outflow properties like their velocity $V_{\rm out}$, their ejected mass rate $\dot M_{\rm out}$ and their loading factor $\eta$ with these main galaxy properties. 

\subsubsection{Scaling relations involving $V_{\rm out}$}
To estimate \Vout, one does not necessarily need a background quasar. 
Indeed, many other studies derived outflow velocities with enough accuracy 
\citep[$\pm 10-20\%$; e.g][]{MartinC_05a, GenzelR_11a, NewmanS_12a, ArribasS_14a, HeckmanT_15a}. 
Those studies found a significant, but scattered, correlation between the outflow velocity and galaxy SFRs at low redshift \citep{HeckmanT_00a, MartinC_05a, RupkeD_05a, MartinC_12a, ArribasS_14a, HeckmanT_15a}. 
\citet{MartinC_05a} derived an upper limit on $V_{\rm out}$ as a function of SFR. 
This limit corresponds to the upper envelope of the outflow velocity distribution at a given SFR. 

We {show} in Appendix A that down-the-barrel and background quasar derived outflow velocities give similar {outflow speeds $V_{\rm out}$} and are therefore comparable.
Also, it is worth mentioning that background quasar measurements are made at larger radii than down-the-barrel and hence suffer from time travel effects that could obscure correlations with SFR if the SFR varies in time. 
This effect is discussed in papers II and III and is one of the reasons only galaxy-quasar pairs that have impact parameters $\leq 100$~kpc were selected in these studies. This reduces the possible effect of this time traveling effect on SFR estimation but does not remove it completely. 

{In Fig.~\ref{fig:vout_vs_sfr}, we investigate the dependence of $V_{\rm out}$ on SFR, SFR surface density, $\Sigma_{\rm SFR}$, and $V_{\rm circ}$, in the
top left, middle and right panels, respectively.}
In the left panel, we corrected { the} SFRs for {the} redshift evolution {of the MS} using \citet{BoogaardL_18} { in order } to have all SFRs at the same redshift ($z=0.2$ in this case to match H15). 
The dashed black line represents the upper outflow velocity found by \citet{MartinC_05a} {for a small sample of local galaxies}.  
They found that $V_{\rm out}$ increases with SFR and their upper limit seems to under-estimate $V_{\rm out}$ for a given SFR.

{The red dot-dashed line in the top left and middle panels represents the} positive correlation found by H15 and \citet{HeckmanT_16a}. 
The red (blue) line {with the shaded area represents our power law fit (which uses least squares fitting method)} of the ``strong'' (``weak'') population, respectively,  { and its error} obtained using the bootstrap method on 100k realizations.
We find that the correlation between \Vout\ and the reduced SFR is positive for the ``strong'' outflows (\Vout$\propto \rm SFR^{0.6\pm0.07}$). 
For the ``weak'' outflows in blue, a {weaker} correlation can be seen (\Vout$\propto \rm SFR^{0.2\pm0.09}$). 
The bootstrap fitting results are given in Table~\ref{table:bootstrap}.

\begin{figure*}
  \centering
  \includegraphics[width=17cm]{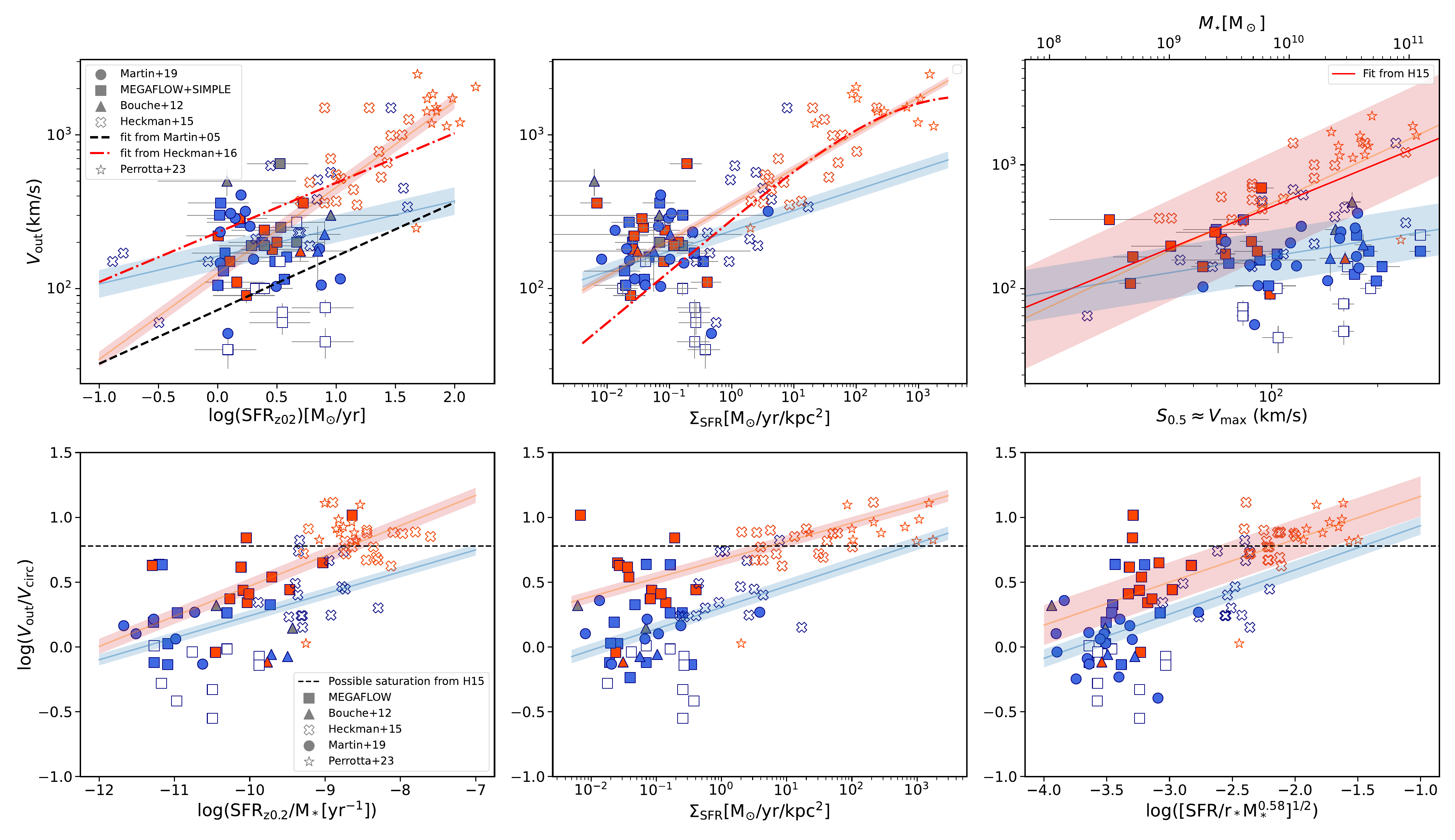}
  \caption{\textit{Top left: }$V_{\rm out}$ as a function of SFR for MEGAFLOW and SIMPLE as well as observations from  {all the samples used in this study}.
  The dashed black line shows a fit ($\log V = (0.35\pm0.06)\log(\rm SFR)+(1.56\pm0.13)$) from \citet{MartinC_05a}. 
  Errors on \citet{HeckmanT_15a} observations are 0.2 dex for SFR and 0.05 dex for $V_{\rm out}$.
  \textit{Top middle: }\Vout\ as a function of $\Sigma_{\rm SFR}$ for the same surveys as in left panel.
  \textit{Top right: }$V_{\rm out}$ as a function of $S_{0.5}$. 
  The red dot-dashed line shows a fit from H16. 
  In these three panels, red (blue) observations correspond to starburst (SFGs) for P23 or ``strong'' (``weak'') outflows for MEGAFLOW, M19 and H15.
  \textit{Bottom : }Outflow velocity normalized by the galaxy circular velocity as a function of the specific SFR (SFR/$M_*$), SFR  surface density ($\Sigma$SFR) and a combination of specific SFR and $\Sigma_{\rm SFR}$ found in H15, from left to right panels.
  }
  \label{fig:vout_vs_sfr}
\end{figure*}

Concerning the correlation between $V_{\rm out}$ and  SFR {surface density} ($\Sigma_{\rm SFR}$), there have been disagreements about its existence \citep[e.g.][]{ChenY_10a, RubinK_14a, GenzelR_11a, NewmanS_12a}. 
In the top middle panel of Figure~\ref{fig:vout_vs_sfr}, we shows the outflow velocity $V_{\rm out}$ as a function of $\Sigma_{\rm SFR}$\footnote{Since $\Sigma_{\rm SFR}$ is correlated by the galaxy size which is correlated with $M_*$, we thus do not normalize for redshift evolution for this quantity.}. 
\citet{HeckmanT_16a} found a strong correlation between those two quantities. 
Adding our observations as well as M19 to their sample confirms { this} correlation. 

We also point out that the H15 $\Sigma_{\rm SFR}$ are large compared to those of MEGAFLOW galaxies. 
Looking at Figure~\ref{fig:ms}, H15 galaxies having larger SFRs than MEGAFLOW ones. To check whether the large $\Sigma_{\rm SFR}$ of H15 were due to only SFRs or also sizes,  
Figure~\ref{fig:r_half} shows the distribution of half-light radius of samples used in this study.
We see that H15 galaxies tend to be a much smaller.
This contributes to the fact that their $\Sigma_{\rm SFR}$ are quite large compared to MEGAFLOW or M19 galaxies.
Using the same bootstrap method as for the first top left panel of this figure, there is no large difference between ``strong'' and ``weak'' outflows, \Vout$\propto \Sigma_{\rm SFR}^{0.2\pm0.03}$ and \Vout$\propto \Sigma_{\rm SFR}^{0.1\pm0.05}$, respectively. 

\begin{figure}
  \centering
  \includegraphics[width=7cm]{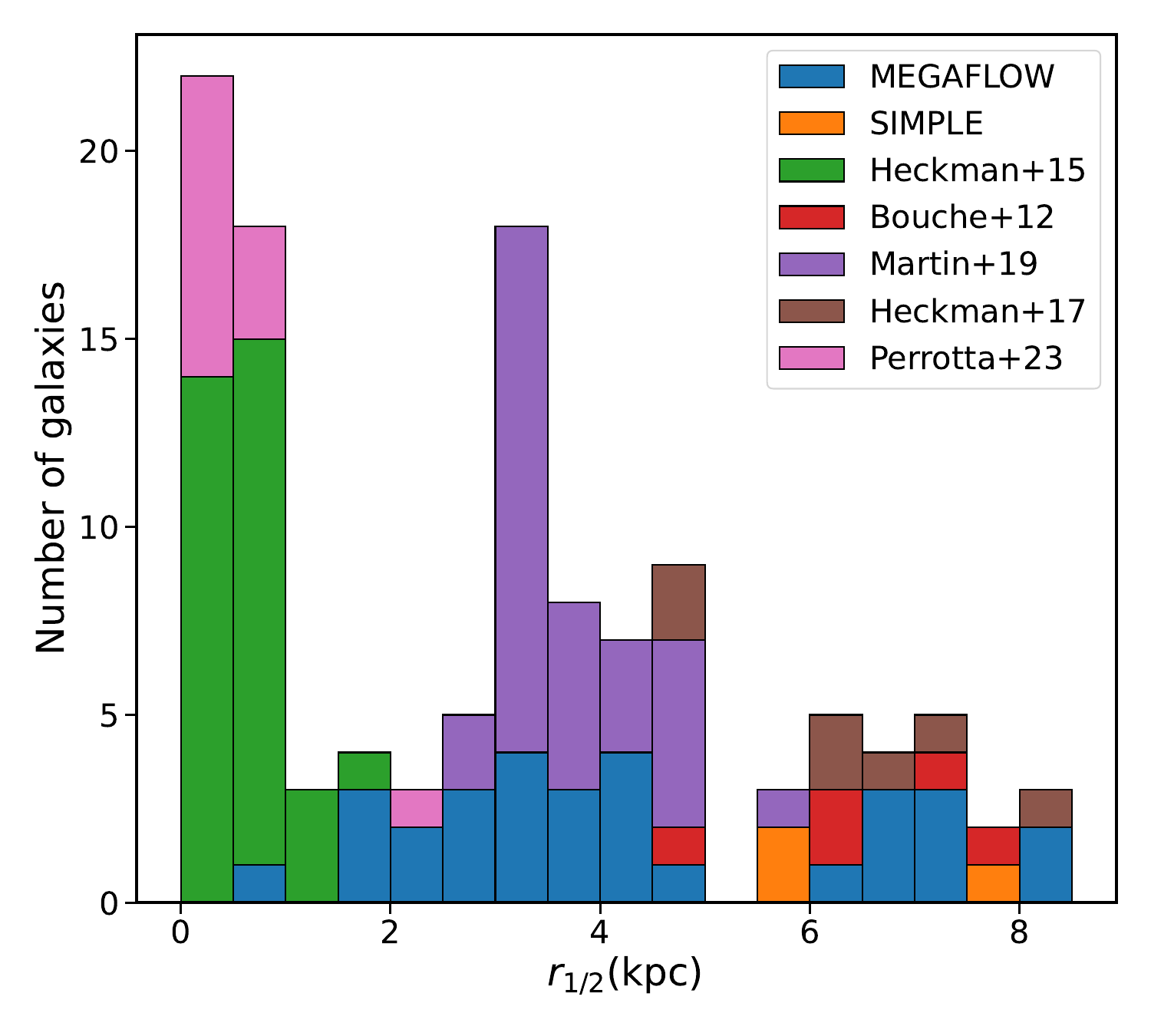}
  \caption{Distributions of half-light radius $r_{1/2}$ for all the samples (MEGAFLOW, SIMPLE, H15, \citet[][]{BoucheN_12a}, M19, \citet[][]{HeckmanT_17a} {and P23} in blue, orange, green, red, purple, maroon and pink, respectively).
  }
  \label{fig:r_half}
\end{figure}

The next step is to investigate whether the outflow velocity depends on the host galaxy mass. 
Following the SFR-$M_\star$ relation (Fig.~\ref{fig:ms}) and the tendency of $V_{\rm out}$ to increase with the host galaxy star formation rate, we could expect the outflow velocity to correlate also with the host galaxy mass. 
However, a more massive galaxy has a deeper gravitational well, and thus it is more difficult for gas to accelerate. 
The top right panel of Figure~\ref{fig:vout_vs_sfr} shows the relation between $V_{\rm out}$ and $S_{0.5}$ 
.
For the ``weak'' outflows from MEGAFLOW and M19 shown in blue, outflow velocities are almost constant at around 100-200~\kms{}
, { while} for ``strong'' outflows (red points) 
 \Vout\ indeed strongly correlates with galaxy mass, which confirms the results from H15, represented by the red curve on this top right panel.
Using the bootstrap fitting method, we indeed find that ``strong'' outflows have a {larger slope} with $S_{0.5}$ (\Vout$\propto S_{0.5}^{1.3\pm0.2}$) than in the case for ``weak'' outflows (\Vout$\propto S_{0.5}^{0.5\pm0.21}$).

To summarize, $V_{\rm out}$ correlates with SFR as well as with $\Sigma_{\rm SFR}$. 
\Vout\ also correlates with the host galaxy mass if outflows are ``strong'' and is { weakly mass-dependent} for ``weak'' outflows. 
In other words, we begin to see a difference in outflow properties between ``weak'' and ``strong'' outflows where ``strong'' outflows appears to correlate more strongly with galaxy properties than ``weak'' ones. 
Distinguishing between those two outflow populations allows us to confirm that the formalism of H15 is relevant to SFGs and starbursts.

In order to  {investigate further}  the  possible scaling relations for the outflow velocity, we show, in
the bottom row of Figure~\ref{fig:vout_vs_sfr}, \Vout\ normalized by the galaxy circular velocity $V_{\rm circ}$ as a function of specific SFR corrected to $z=0.2$ (bottom left panel), SFR surface density ($\Sigma_{\rm SFR}$) (bottom middle panel) and a combination of specific SFR and $\Sigma_{\rm SFR}$ (bottom right panel). 

{One sees that the relative outflow speed $\Vout/V_{\rm circ}$ is a simple function (universal?) of SFR, or momentum injection rate (bottom left), and } that, as mentioned in H15, there is  a possible saturation in normalized outflow velocity when $\Vout \approx 4 V_{\rm circ}$ above $\Sigma_{\rm SFR}\sim 10 ~\rm M_\odot \rm yr^{-1} \rm kpc^{-2}$.
This saturation is shown with the horizontal black dashed line in each panel of the bottom row of Fig~\ref{fig:vout_vs_sfr}.
We can see that this saturation looks to be the case if we only look at H15 data. However, with the addition of the lower SFR data (MEGAFLOW + M19), that is no longer as convincing. 
In these bottom panels we also show the bootstrap fits for each outflow population in red and blue lines for ``strong'' and ``weak'', respectively.
Apart from the normalization of each fit, ``strong'' and ``weak'' outflows appear to correlate similarly with each quantity. 
Those correlations are less scattered than with SFR, $\Sigma_{\rm SFR}$ or $M_\star$ with which we can directly compare with the top row and there is no clear differentiation between both outflow populations if we consider them together or independently (as shown by the corresponding correlation coefficients in Table~\ref{table:bootstrap}).

\subsubsection{Scaling relations for $\dot M_{\rm out}$}

We now turn to another fundamental wind property, namely the ejected mass outflow rate $\dot M_{\rm out}$ (and the mass loading factor $\eta\equiv \dot M_{\rm out}/$SFR) 
{and investigate possible scaling relations with the properties of the host galaxies.}

{In order to compare the outflow ejection rates $\dot M_{\rm out}$ of paper~III to H15 who estimated the mass ejection rate assuming spherically symmetric outflows with $\thetam=4\pi$sr }
, we scaled their $M_{\rm out}$  to 60$^\circ$ 
using, as previously, \Vout\ from \citet{HeckmanT_16a}. 

Concerning M19 galaxies, to estimate the mass ejection rate we used the \MgII\ REW as a proxy  to estimate the $N_{\rm H}$ column density \citep{MenardB_09a, ZhuG_13a}.
This proxy is only viable for {strong} \MgII\ REW and we thus only select the galaxies with $W_r^{\lambda 2796}>0.5\AA$.

Similar to Figure~\ref{fig:vout_vs_sfr}, 
{the top panels of}  Figure~\ref{fig:mout_vs_sfr} shows the mass ejection rate ($\dot M_{\rm out}$) as a function of SFR corrected to $z=0.2$ (left panel), $\Sigma_{\rm SFR}$ (middle panel) and the galaxy  mass (right panel). 
{Regarding the $\dot M_{\rm out}$-SFR relation,
}
\citet{HopkinsP_12a} predicted that $\dot M_{\rm out}\propto \rm SFR^{0.7}$ (shown as the dashed red line in the left panel) whereas \citet{ArribasS_14a} observed a steeper index $\dot M_{\rm out}\propto \rm SFR^{1.11}$ (shown as a solid black line).

The amount of ejected mass by supernova explosions being directly linked {to} SFR, it is intuitively expected that SFR correlates with $\dot M_{\rm out}$. 
Looking {by eye} only at ``weak'' outflows in blue, there is no obvious correlation, but for ``strong'' outflows (in red) the correlation between the mass ejection rate and the galaxy SFR appears to be more in agreement with \citet[][]{HopkinsP_12a}'s predictions than with the observations of \citet{ArribasS_14a}. 
As for \Vout, we use the bootstrap fitting method to measure the relations between $\dot M_{\rm out}$ and the galaxy properties. 
Looking at the top left panel of Figure~\ref{fig:mout_vs_sfr}, both ``strong'' and ``weak'' outflows correlate  slightly with SFR and bootstrap slopes are close to each other{, with the ``strong'' outflows having a {steeper slope} than the ``weak'', $\dot M_{\rm out} \propto \rm SFR^{0.5\pm0.1}$ and $\propto \rm SFR^{0.3\pm0.2}$} for both {``strong'' and ``weak''} populations{, respectively}. 

\begin{figure*}
  \centering
  \includegraphics[width=17cm]{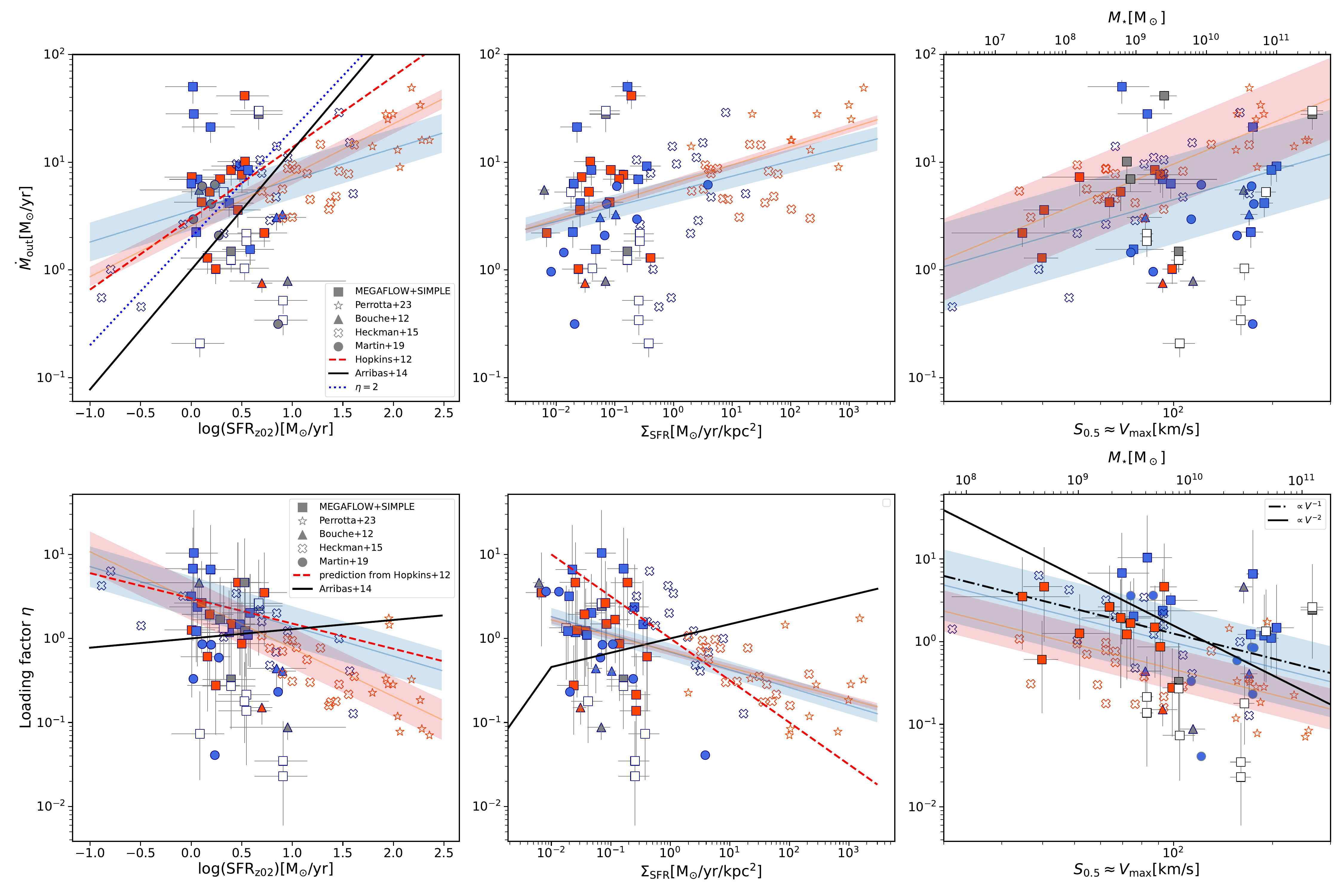}
  \caption{\textit{Top: }Mass ejection rate as a function of star formation rate (left), star formation rate surface density ($\Sigma_{\rm SFR}$, middle), and the galaxy circular velocity (right, and stellar mass on top x-axis) for both surveys (MEGAFLOW and SIMPLE) as well as observations from B12, H15, M19 {and P23}. 
  In the left panel, the dashed red line shows the prediction $\dot M_{\rm out}\propto \rm SFR^{0.7}$ from \citet{HopkinsP_12a} and the black line shows $\dot M_{\rm out}\propto \rm SFR^{1.11}$ observed by \citet{ArribasS_14a}. 
  The blue dotted line corresponds to a constant mass loading factor $\dot M_{\rm out}/$SFR of 2.
  Errors for \citet{HeckmanT_15a} are 0.25 dex for $\dot M_{\rm out}$ and 0.2 dex for SFR and $\Sigma_{\rm SFR}$. 
  \textit{Bottom: }$\eta$ as a function of SFR (left), $\Sigma_{\rm SFR}$ (middle) and $S_{0.5}$ (right) for both surveys (MEGAFLOW and SIMPLE) as well as observations from 
  B12, H15, M19 {and P23}. 
  In the left panel, the dashed red line shows the prediction $\eta \propto \rm SFR^{-0.3}$ from \citet{HopkinsP_12a}  
  and the black line shows the fit $\eta \propto \rm SFR^{0.11}$ from \citet{ArribasS_14a}. 
  In the middle panel, the dashed red line shows the prediction $\eta \propto \Sigma_{\rm SFR}^{-1/2}$ from \citet{HopkinsP_12a} 
  and the black line shows the fit $\eta \propto \Sigma_{\rm SFR}^{0.17}$ from \citet{ArribasS_14a}. 
  Again, errors for \citet{HeckmanT_15a} are 0.2 dex for SFR (and $\Sigma_{\rm SFR}$) and 0.45 dex for $\eta$. 
  In the right panel, the dashed dotted black line shows $\eta \propto V^{-1}$ and the solid black line shows $\eta \propto V^{-2}$.
  }
  \label{fig:mout_vs_sfr}
\end{figure*}

{Contrary to} the correlation between $V_{\rm out}$ and $\Sigma_{\rm SFR}$, the top middle panel of Fig.~\ref{fig:mout_vs_sfr} { shows that} there is no correlation between the mass {outflow} rate and $\Sigma_{\rm SFR}$,
{except perhaps a mild relation, which is confirmed by the bootstrap fitting as $\dot M_{\rm out} \propto \Sigma_{\rm SFR}^{0.16\pm0.04}$ for ``strong'' and $\propto \Sigma_{\rm SFR}^{0.13\pm0.1}$ for ``weak'' outflows. }
Similarly, 
the top right panel of Fig.~\ref{fig:mout_vs_sfr} 
 indicates a weak correlation between the ejected mass rate and $S_{0.5}$ (and thus its stellar mass) for both ``strong'' and ``weak'' outflows, albeit also with a large scatter ($\dot M_{\rm out} \propto S_{0.5}^{0.8 \pm0.4}$ for {the ``weak''} populations {and $\dot M_{\rm out} \propto S_{0.5}^{1.3 \pm0.2}$} for the ``strong'' ones).

\subsubsection{Scaling relations for $\eta$}
\label{subsection:scaling_eta}

The last but maybe the most important parameter concerning galactic outflows is the mass loading factor $\eta$, i.e. the ratio between the mass ejection rate $\dot M_{\rm out}$ and the SFR of the galaxy;
\begin{equation}
 \eta=\frac{\dot M_{\rm out}}{\rm SFR} \propto \rm SFR^{\alpha-1} \label{eq:eta_sfr}
\end{equation}
if $\dot M_{\rm out} \propto \rm SFR^\alpha.$

Depending on the value of $\alpha$ in Equation~\ref{eq:eta_sfr}, we can differentiate 3 cases: 
the correlation between $\eta$ and SFR is either positive ($\alpha>1$), 
negative ($\alpha<1$) or $\eta$ can be constant ($\alpha=1$). 
Those three possibilities are represented by the lines in top left panel of Figure~\ref{fig:mout_vs_sfr} with $\alpha=0.7$ \citep{HopkinsP_12a}, $\alpha=1.11$ \citep{ArribasS_14a} and $\alpha=1$ (a constant loading factor $\eta=2$), 
none of which actually fit the data. 
If a correlation exists, it has a lower $\alpha$ than \citet{HopkinsP_12a}. 
Indeed, according to the result of the bootstrap method in Table~\ref{table:bootstrap}, $\alpha$ is around {0.4} regardless of outflow strength. 
Thus, we can argue that if $\eta$ correlates with SFR, it should be an anti-correlation ($\alpha < 1$). 
The bottom left panel of Figure~\ref{fig:mout_vs_sfr} shows the mass loading factor as a function of SFR. 
We can see a scattered anti-correlation between these two properties. 
This confirms that $\eta \propto \rm SFR^{\alpha-1}$ with $\alpha<1$. 
Our observations are thus in qualitative agreement with \citet{HopkinsP_12a} predictions. 

As there is no significant difference between ``weak'' and ``strong'' outflows we can conclude that the mass loading factor indeed anti-correlates with the galaxy SFR regardless of the galaxy type.
Thus, galaxies with high SFR tend to have a lower mass loading factor than galaxies with a lower SFR. 
{We will return to the implications of this result in \S~\ref{section:samples_mechanisms}.}

{As before, we would like to investigate}
whether the mass loading factor depends on local galaxy properties (such as $\Sigma_{\rm SFR}$) or global (such as mass). 
In the bottom middle panel of Figure~\ref{fig:mout_vs_sfr}, we show $\eta$ as a function of $\Sigma_{\rm SFR}$. 
Prediction and observations from \citet{HopkinsP_12a} and \citet{ArribasS_14a} are represented by red-dashed and black lines respectively. 
It appears that there is a clear anti-correlation between $\eta$ and $\Sigma_{\rm SFR}$ in the data. 
The slope of this anti-correlation seems to be the same for ``strong'' and ``weak'' outflows and the bootstrap fitting gives us the same slope for both populations ($\approx$-0.2).

It is worth mentioning that \citet{ArribasS_14a} and \citet[][]{NewmanS_12a} found that galaxies (low-$z$ galaxies from \citet{ArribasS_14a} and high-$z$ galaxies from \citet{NewmanS_12a}) require a $\Sigma_{\rm SFR}$ larger than 1 \sigmpy\ for launching strong outflows. 
This statement is not supported by either MEGAFLOW or M19 galaxies since the majority of them show outflow signatures and have $\Sigma_{\rm SFR}$ below 1 \sigmpy, even if we only focus on ``strong'' outflows shown in red. 

As seen in bottom right panel of Figure~\ref{fig:mout_vs_sfr}, the mass loading factor anti-correlates with the galaxy stellar mass. 
Concerning the correlation slopes, we will discuss their implication in the next section. 

Another aspect about the mass loading factor is its redshift dependence. 
Indeed, as there is a peak in star-formation density at redshift 2-3 \citep[e.g. ][]{LillyS_96a, MadauP_96a, BehrooziP_13a}, if $\eta$ correlates with SFR, one can expect a correlation between $\eta$ and redshift. 
We thus investigated  this relation but found no evident correlation \citep[as compared to][]{MuratovA_15a}.
Figure~\ref{fig:eta_redshift} in the appendix shows $\eta$ as a function of redshift for individual cases of each study and shows no apparent correlation. 
{Finally, we find no correlation between $\eta$ and \Vout{}, in agreement with results from H15.
}

\section{What mechanisms drive galactic winds?}
\label{section:samples_mechanisms}

We will now {use the results shown in the previous section to tackle the} question what mechanisms drive galactic winds out of the galactic disk.
To date, there are two major mechanisms which could be responsible for driving materials out of the galaxy: energy-driven and momentum-driven outflows
\citep[as reviewed in][]{HeckmanT_17a}.
The momentum-driven wind scenario considers that the two primary sources of momentum deposition in driving galactic winds are supernovae and radiation pressure from the central starburst. 
{This model assumes that $\dot{M}_{\rm out}\Vout$ is constant and  implies that $\eta$ must be inversely proportional to $V_{\rm circ}$, i.e. $\eta \propto V_{\rm circ}^{-1}$, given that} \Vout\ scales as $S_{0.5}$ and thus as the galaxy circular velocity \citep[e.g.][]{MartinC_05a, OppenheimerB_06a,OppenheimerB_08a, DaveR_11b,HeckmanT_17a}.

Energy-driven wind model assumes that when stars evolve, they deposit energy into the ISM. 
The amount of gas blown out of the disk is assumed to be proportional to the total energy released by supernovae and inversely proportional to the escape velocity squared. 
In the energy-driven scenario, energy conservation implies $\eta \propto V_{\rm circ}^{-2}$ \citep[e.g.][]{vandenBoschF_02a}.

In the bottom right panel of Figure~\ref{fig:mout_vs_sfr}, we show the mass loading factor $\eta$ as a function of galaxy  $S_{0.5}$\footnote{We choose to use $S_{0.5}$ as we used this parameter to derive galaxy stellar masses. It is more appropriate to use this factor than the maximum rotation velocity as some of our galaxies are dispersion-dominated.} (bottom x-axis) and galaxy stellar mass (top x-axis). 
In addition, we also show $\eta \propto V^{-2}$ (the black line in the bottom right panel) in order to see if we could discriminate between the two mechanisms for driving outflows. 

\begin{figure}
  \centering
  \includegraphics[width=8cm]{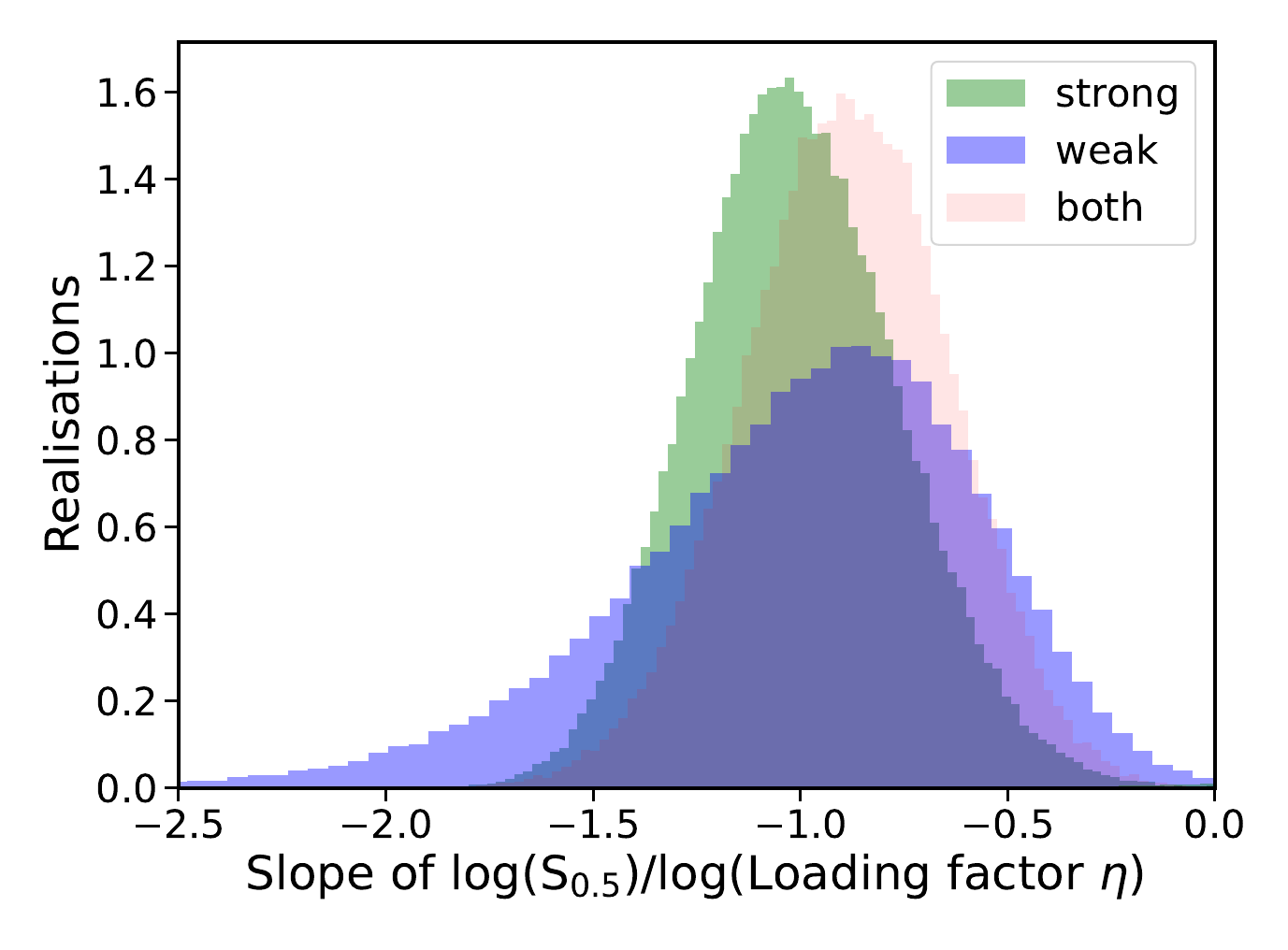}
  \caption{Histogram of fitted slopes for 100k bootstrap {normalized} realisations of the ``weak'' outflows (blue), ``strong'' outflows (green) and the combined samples (red).}
  \label{fig:eta_vs_s05_slope}
\end{figure}

If we do not distinguish between ``weak'' and ``strong'' outflows, we find a {scatter} anti-correlation between $\eta$ and $S_{0.5}$ with a slope of $\approx -0.9\pm0.3$. 
This slope value does not allow us to differentiate between momentum and energy-driven scenarios {but points toward a momentum-driven scenario nonetheless}.
Looking at ``strong'' and ``weak'' outflow populations individually, we find that the $\eta$ anti-correlation with the galaxy stellar mass is steeper. 
{Applying the bootstrap method to create 100k realisations of the ``weak'' and ``strong'' groups with 39 and {43} data points respectively, as well as for the combined data, the resulting histogram of the fitted slopes of the loading factor $\eta$ -- $S_{0.5}$ relation is shown in Figure~\ref{fig:eta_vs_s05_slope}. 
We can see that using all the data points, the mass loading factor $\eta \propto V^{-0.89 \pm 0.25}$ whereas for each population independently, we find $\eta \propto V^{-0.99 \pm 0.43}$ {and $\eta \propto V^{-1.01 \pm 0.24}$  for ``weak'' and ``strong'' outflows, respectively}.

This slope is consistent with the prediction of \citet{HopkinsP_12a} ($\eta \propto S_{0.5}^{-1.2\pm0.2}$) and favors a momentum-driven scenario for galactic outflows.

\section{Summary and Conclusions}
In this paper, we used the results published in Paper III on outflow properties inferred from quasar absorption lines and compared them with other studies reporting mass ejection rates in order to investigate possible scaling relations between outflows and their host galaxy properties. 
The three main parameters we investigated are the outflow velocity \Vout, the mass ejection rate $\dot M_{\rm out}$ and the mass loading factor $\eta$. 
Those parameters were related to global galaxy properties like SFR, stellar mass and SFR surface density.

We distinguished between two outflow regimes: ``weak'' and ``strong'' outflows (see \S~\ref{subsection:previous_studies}).
These regimes are the two cases where their momentum flux is larger or lower than {ten times} the critical momentum flux required to have a net outward force on an outflowing cloud, i.e. ``weak'' if $\log(\dot p_*/\dot p_{crit})<1.0$ and ``strong'' if $\log(\dot p_*/\dot p_{crit})>1.0$.
For each parameter combination, we used a bootstrap method in order to estimate the power law slopes of the relations between the aforementioned properties. 
The two regimes show different behaviours as can be summarize as follows:

\begin{table}
\centering
\caption{Summary of bootstrap fitting}
\label{table:bootstrap}
\begin{tabular}{lccc}
\hline
Parameters & ``strong''    & ``weak''  & Both  \\

(1)   &  (2)          & (3)               & (4)  \\
\hline
{$\log(\Vout)$}   &  &  &  \\
 $\log(\rm SFR_{z0.2})$  &  {0.56$\pm$0.05}   &  0.18$\pm$0.09   &  {0.46$\pm$0.05}\\
 $\log(\Sigma_{\rm SFR})$&  {0.23$\pm$0.03}   &  0.13$\pm$0.05   &  {0.23$\pm$0.02}\\
 $\log(S_{0.5})$         &  {1.33$\pm$0.20}   &  0.45$\pm$0.21   &  {0.77$\pm$0.18}\\
\hline
$\log(\dot M_{\rm out})$ &  &  &  \\
 $\log(\rm SFR_{z0.2})$  &  {0.47$\pm$0.09}   &  0.29$\pm$0.17   &  {0.40$\pm$0.08}\\
 $\log(\Sigma_{\rm SFR})$&  {0.16$\pm$0.04}   &  0.13$\pm$0.11   &  {0.17$\pm$0.04}\\
 $\log(S_{0.5})$         &  {1.27$\pm$0.21}   &  0.89$\pm$0.41   &  {1.09$\pm$0.21}\\
 \hline
 $\log(\eta)$ &  &  &  \\
 $\log(\rm SFR_{z0.2})$  &  {-0.57$\pm$0.24}   &  {-0.35$\pm$0.24}   &  {-0.43$\pm$0.13}\\
 $\log(\Sigma_{\rm SFR})$&  {-0.18$\pm$0.04}   &  {-0.21$\pm$0.10}   &  {-0.19$\pm$0.04}\\
 $\log(S_{0.5})$         &  {-1.01$\pm$0.24}   &  {-0.99$\pm$0.43}   &  {-0.89$\pm$0.25}\\
 \hline
\end{tabular}\\
{
(1) Wind parameters: outflow velocity \Vout, ejected mass rate $\dot M_{\rm out}$ and the mass loading factor $\eta$ as a function of (if any correlation) $\rm SFR_{z0.2}$, $\Sigma_{\rm SFR}$ and $S_{0.5}$;
(2) ``strong'' outflow population;
(3) ``weak'' outflow population;
(4) Both populations altogether.
}
\end{table}
The outflow velocity correlates with SFR, $\Sigma_{\rm SFR}$ and $S_{0.5}$ and shows stronger correlations for the ``strong'' outflow population. 
In particular, \Vout\ exceeds the upper limit of \citet{MartinC_05a} concerning its correlation with SFR.

The mass ejection rate $\dot M_{\rm out}$ correlates, as the outflow velocity, with the three galaxy properties for both populations but the ``strong'' outflows does not {clearly} correlate with SFR surface density.

Finally, the mass loading factor anti-correlates with SFR, $\Sigma_{\rm SFR}$ and $S_{0.5}$ for both populations. 
However, $\eta$ is apparently not redshift dependent. 
Details on the different slopes are summarized in Table~\ref{table:bootstrap}.

We also find that the galaxy does not need $\Sigma_{\rm SFR}>1$~\mpy~kpc$^{-2}$ in order to be able to launch material out of the galactic disk.

In addition, we addressed the question which mechanism is dominant and$/$or responsible for launching outflows.    
According to the bottom right panel of Figure~\ref{fig:mout_vs_sfr}, we find that both ``weak'' and ``strong'' outflows point towards a momentum-driven scenario as the coefficient found for both populations is close to $\eta \propto V^{-1}$ with $\eta \propto V^{-1.01\pm0.24}$ {and $\eta \propto V^{-0.99\pm0.43}$}. 
This result needs to be confirmed with additional and more accurate results but it shows that depending on the outflow ``strength'', the mechanism responsible for launching the gas tends to be the same for the ``strong'' and ``weak'' outflow regimes.   

In conclusion, using a bootstrap method on all galaxies and for the two regimes individually, we saw that one needs to differentiate between ``strong'' and ``weak'' outflows as both regimes have different behaviors. 
This differentiation is thus important to understand the role of galactic outflows in galaxy formation and evolution. 
We compared outflow properties derived from quasar absorption line and down the barrel methods and showed that a universal formalism can be used for outflows regardless of the method used. 
We mentioned that the background quasar line method has larger impact parameter than down the barrel and can suffer from time travel effects that could obscure correlation with SFR if the SFR varies during the time needed for the gas to get from the galaxy to the quasar line of sight. 
As this effect is discussed in previous papers (papers I and III), we do not develop this effect here but are aware that this may have an effect on results implying SFRs.
Some results on properties like mass loading factors or mass ejection rates have order of magnitudes uncertainties and are more indicative than accurate but allowed us to nonetheless draw some conclusions using a bootstrap fitting method. 
Using the MEGAFLOW results on outflow properties and differentiating between ``weak'' and ``strong'' outflows, we confirm scaling relations as well as open new paths in the understanding of galactic winds properties and thus the evolution and formation of galaxies. 
Accuracy is essential in order to get a correct answer for those scaling relations, especially concerning wind properties like the mass outflow rate and mass loading factor. 
The background source method would greatly benefit from an accurate estimation of the hydrogen column density to be able to estimate lower column densities {that} can be inferred from \MgII\ absorption. 
Therefore, future observations are still needed. The James Webb Space Telescope allows for higher redshift outflow studies and will provide {many} more outflow cases.

\section*{Acknowledgments}
{We thank the referee for her/his helpful comments and suggestions which helped to greatly improve the paper.}
This work has been carried out thanks to the support of the Agence Nationale de la Recherche (ANR) grant 3DGasFlows (ANR-17-CE31-0017) {as well as support from the Centre National d'Etudes Spatiales (CNES) through the APR program}.

\bibliographystyle{aa}
\bibliography{references}
\label{lastpage}

\section*{Appendix}

\subsection*{On measuring outflows speeds}

To estimate the outflow velocity, there are differences between background quasar and galaxy absorption (aka ``down the barrel'') methods. 
The main difference is the background object. 
For quasar sightlines, it is known that the probed gas is likely to be in the CGM while for galaxies (down the barrel), the gas can be anywhere in the CGM or IGM towards the observer.
A background quasar also gives the location of the absorbing gas, namely the impact parameter, whereas absorption in a galaxy spectrum does not provide such information and is usually assumed to be several kpc from the host galaxy.

In addition to this difference, the outflow absorption profile is different. 
In H15, the observer looks directly at the galaxy. 
The outflowing gas ejected from this galaxy is moving toward the observer. 
Thus, this gas gives rise to blue-shifted absorption in the galaxy spectrum. 
In order to see this blue-shifted absorption, the host galaxy needs to have a low inclination (to be close to a face-on configuration). 
Using a background quasar, the outflow absorption can be either blue or red-shifted with respect to the host galaxy systemic redshift. 
In addition, host galaxies are selected to be not face-on.  
As a matter of fact, host galaxies selected in paper III needed to have an inclination $i\geq35^\circ$ for low position angle uncertainties. 

To assess those differences and thus confirm that we obtain similar results on the outflow velocity using our wind model, we first need to create a configuration similar to the H15 method, namely a down-the-barrel configuration.
We then create a wind model for this specific geometry.
For a face-on galaxy, H15 use the outflow velocity value corresponding to $80-90\%$ of the blue-shifted absorption produced by the outflowing gas. 
This value will give the outflow velocity \Vout$_{,90}$.
This \Vout$_{,90}$ is then corrected in \citet{HeckmanT_16a} to have the maximum outflow velocity of the gas. 
This maximum outflow velocity corresponds to our definition of \Vout.
We thus try to see if the \Vout\ derived by H15 is similar to the one we derive from our wind model. 

The aim is to see if we can reproduce the blue-shifted absorption shape of their data seen in Figure 1 of their paper and also where \Vout\ ends up. 
We create a wind model of a galaxy with an inclination of $i=0^\circ$, azimuthal angle of 90$^\circ$ and an impact parameter $b$ of 0 kpc. 

This model is shown in Figure~\ref{fig:vout_heckman}. 
The top left panel of this Figure is a representation of the sky plane of the face-on galaxy with the outflowing cone directed towards the observer. 
The top right panel represents a side view of the system, showing the line of sight (LOS) in orange crossing the outflowing cone from right to left. 
Since the LOS crosses all the way from the galaxy to the outer part of the cone, we create an accelerated wind model \citep[as we are tracing the accelerating part of the outflow, this model is described in][]{SchroetterI_15a}). 
This accelerated wind model changes the asymmetry of the profile as there are more clouds with lower velocity close to the galaxy.

The bottom panel of Figure~\ref{fig:vout_heckman} shows the resulting absorption profile of this configuration. 
The red vertical dashed line represents the input \Vout. 
We see that this outflowing velocity corresponds to the furthest part of the blue-shifted absorption. 
This is in agreement with the \Vout\ derived by H15, corrected in \citet{HeckmanT_16a}.
We can thus directly compare our results with those of H15. 

\begin{figure}
  \centering
  \includegraphics[width=8cm]{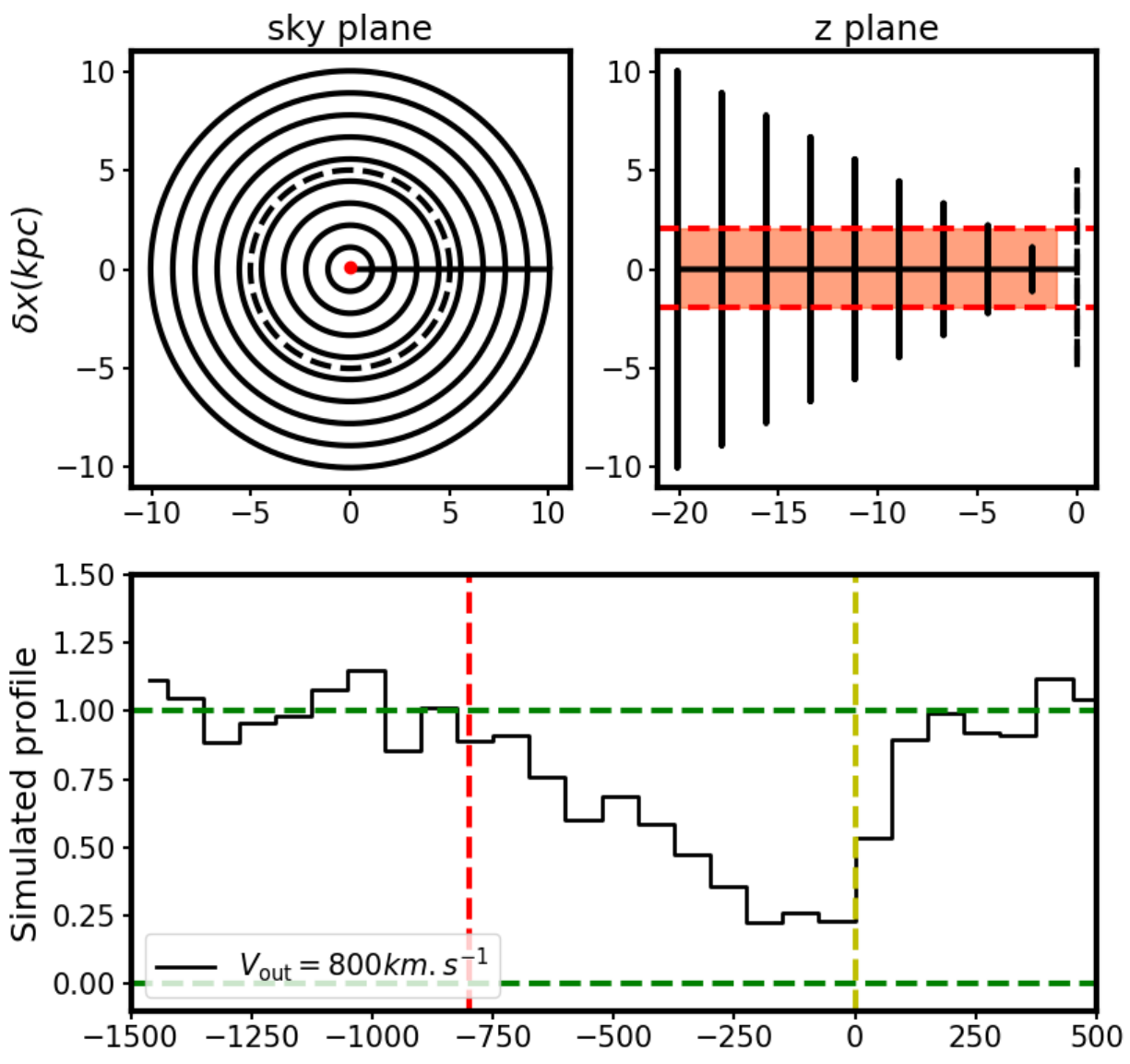}
  \caption{Wind model using a similar configuration as in H15 paper. 
  \textit{Top left: }sky plane representation of the system. the dashed black circle corresponds to the face-on galaxy.
  The black circles represent the outflowing cone and the red dot represents the center of the LOS crossing the outflowing cone.
  \textit{Top right: }side view of the system. 
  the galaxy is represented on the right part in dashed vertical line, the outflowing cone by the vertical black lines going to the left. 
  The line of sight is represented by the red-filled rectangle crossing the outflowing cone.
  \textit{Bottom: }the resulting simulated absorption profile. 
  The vertical red dashed line represents the outflow velocity used as an input for this wind model. 
  The simulated profile has been convolved with the resolution used in H15 paper ($\sim75$\kms\ FWHM).
    }
  \label{fig:vout_heckman}
\end{figure}

Even if we do not include galaxies from \citet{ArribasS_14a} 
we will still consider the relations they found to see if there are significant differences between SFGs and Ultra/Luminous infrared galaxies outflow properties.

Concerning M19 galaxies, they use the background quasars method and thus we can easily derive their outflow velocities. 
For their galaxies, we use the maximum velocity offsets (blue or red-shifted) of the \MgII\ absorptions seen in background quasars as projected outflow velocities. 
Then, using the inclination derived in their study and a cone opening angle of 30$^\circ$, we get the estimated outflow velocities $V_{\rm out}$. 
From those outflow velocities, impact parameters and $W_r^{\lambda 2796}$, we estimate the mass outflow rates $\dot M_{\rm out}$ using equation 5 of \citet[][]{SchroetterI_15a}. 
Then, mass loading factors $\eta$ are estimated using SFRs. 
Since their SFRs are derived using M19 main sequence figure, we emphasize that those results are more indicative than accurate.

{For P23 outflow velocities, like M19, we assume the outflow velocity to be the maximum \MgII\ absorption velocity offsets. We note that they also have \FeII\ absorption velocities but for consistency we choose to only use the \MgII{} ones since we do the same for background quasars. P23 also already have ejected mass outflow rate for bi-conical outflow geometry as well as corresponding loading factors, we thus do not need to re-estimate them. }

\subsection*{Mass loading factor redshift evolution}
Figure~\ref{fig:eta_redshift} shows the mass loading factor as a function of host galaxy redshift. 
As mentioned in the text, there is no apparent correlation between the mass loading factor and the host galaxy redshift.
\begin{figure}
  \centering
  \includegraphics[width=8cm]{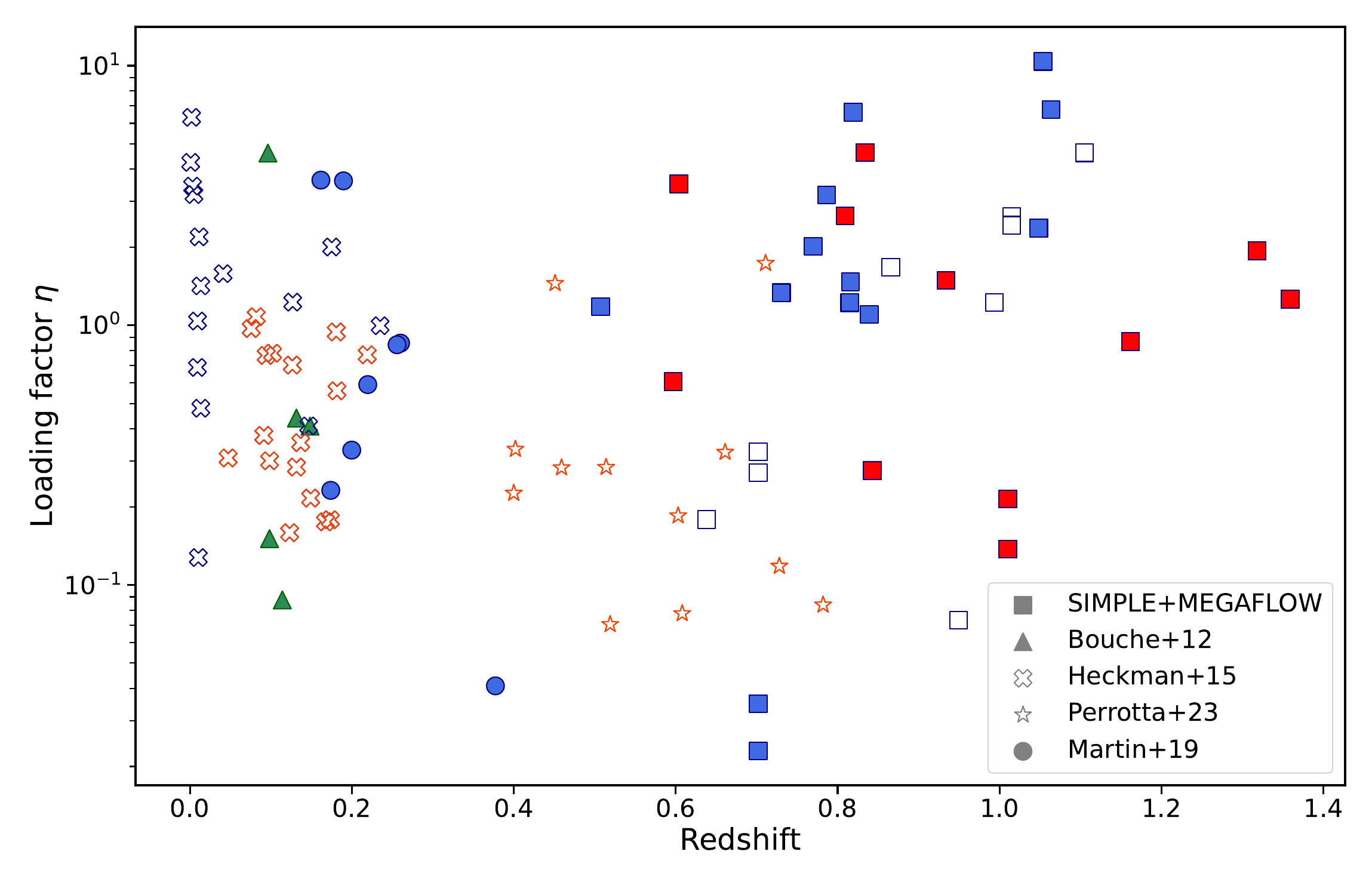}
  \caption{Mass loading factor as a function of host galaxy redshift.
  All points are individual results of both $\eta$ and redshift from the studies mentioned in the legend. 
    }
  \label{fig:eta_redshift}
\end{figure}

\subsection*{Outflow velocity versus impact parameter}
Figure~\ref{fig:vout_b} shows the outflow velocity as a function of impact parameter for background source studies. 
As mentioned in the text, there is no apparent correlation between those parameters. 

\begin{figure}
  \centering
  \includegraphics[width=8cm]{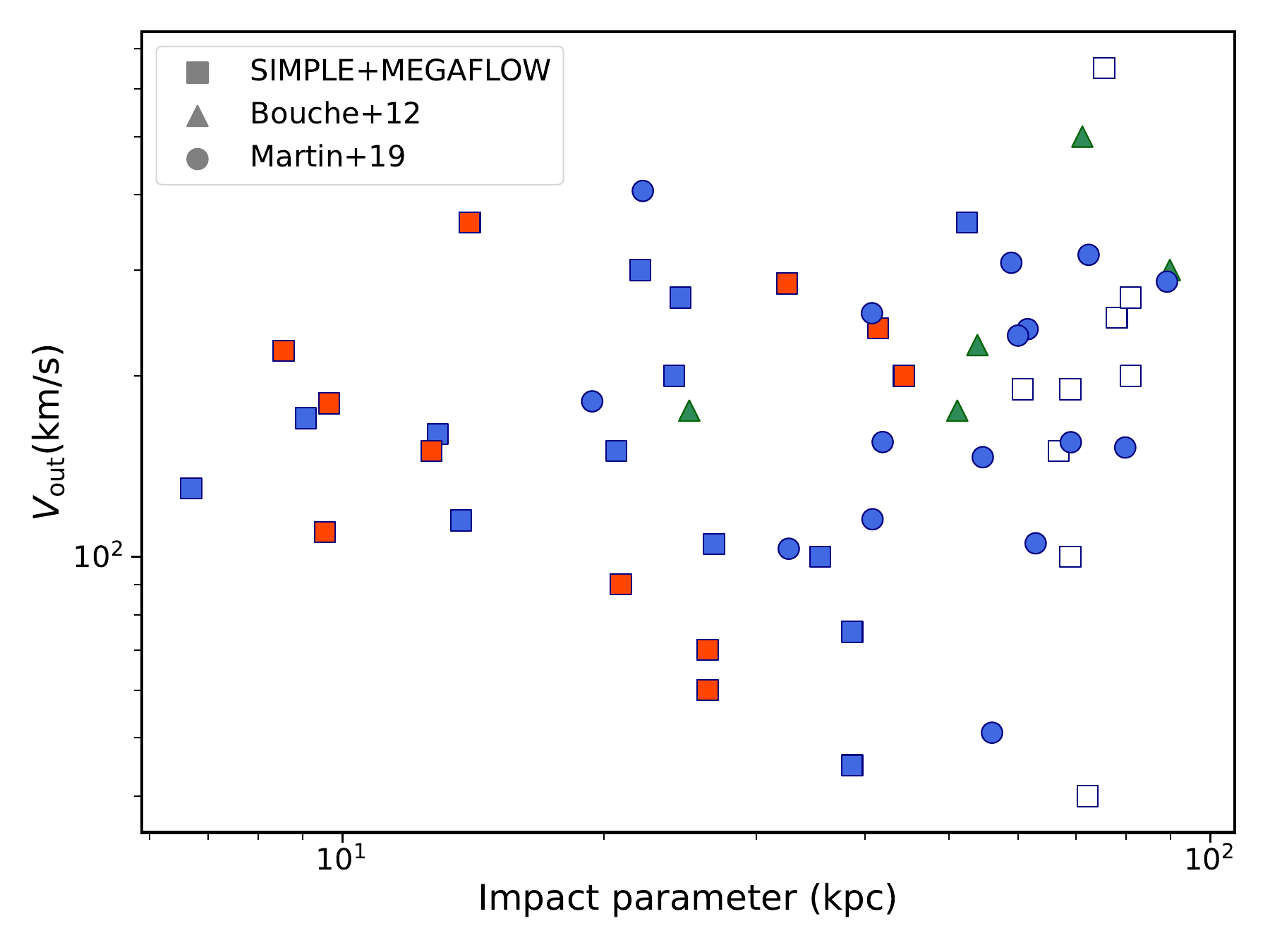}
  \caption{Outflow velocity as a function of the impact parameter for background sources studies. 
    }
  \label{fig:vout_b}
\end{figure}

\subsection*{Galaxy properties}
The following tables list all the galaxy properties used in this paper in order to be able to reproduce the results.

\begin{table*}
\centering
\caption{Summary of galaxy properties}
\label{table:gal_prop}
\begin{tabular}{lccccccccc}
\hline
Study & V$_{\rm out}$ & $\log{(\rm SFR_{0.2})}$ & $\Sigma_{\rm SFR}$ & $S_{0.5}$ & $\dot M_{\rm out}$ & $\eta$ & $\dot p_\star$ & $\log(\dot p_{\rm crit, c})$ & $\log(\dot p_{\rm crit, s})$ \\
(1) & (2) & (3) & (4) & (5) & (6) & (7) & (8) & (9) & (10) \\
\hline
Paper III  & 180.0 & 0.461 & 0.0253 & 40.38 & 1.799 & 4.984 & 33.53 & 32.35 & 32.60 \\
$\cdots$ & 360.0 & 0.025 & 0.0698 & 83.16 & 14.04 & 10.39 & 34.11 & 33.75 & 33.85 \\
$\cdots$ & 200.0 & 0.376 & 0.0254 & 188.6 & 2.085 & 1.179 & 34.22 & 33.84 & 35.27 \\
$\cdots$ & 100.0 & 0.320 & 0.0177 & 190.8 & 2.642 & 1.334 & 34.27 & 34.22 & 35.29 \\
$\cdots$ & 150.0 & 0.482 & 0.3487 & 205.5 & 4.593 & 1.470 & 34.47 & 34.16 & 35.42 \\
$\cdots$ & 170.0 & 0.062 & 0.2500 & 92.67 & 3.464 & 2.367 & 34.14 & 33.26 & 34.04 \\
$\cdots$ & 70.0 & 0.546 & 0.2656 & 82.87 & 1.079 & 0.249 & 34.61 & 32.93 & 33.84 \\
$\cdots$ & 650.0 & 0.527 & 0.1906 & 93.61 & 20.69 & 4.620 & 34.63 & 33.50 & 34.06 \\
$\cdots$ & 160.0 & 0.583 & 0.0470 & 75.55 & 0.776 & 3.025 & 33.39 & 32.58 & 33.68 \\
$\cdots$ & 90.0 & 0.242 & 0.0237 & 98.98 & 0.509 & 0.275 & 34.24 & 32.85 & 34.15 \\
$\cdots$ & 250.0 & 0.533 & 0.0379 & 72.09 & 5.061 & 1.226 & 34.59 & 33.14 & 33.60 \\
$\cdots$ & 100.0 & 0.394 & 0.1627 & 103.6 & 0.617 & 0.270 & 34.33 & 33.85 & 34.23 \\
$\cdots$ & 45.0 & 0.908 & 0.2547 & 160.0 & 0.170 & 0.022 & 34.85 & 33.98 & 34.99 \\
$\cdots$ & 240.0 & 0.394 & 0.0848 & 87.68 & 4.236 & 1.489 & 34.43 & 33.37 & 33.94 \\
$\cdots$ & 270.0 & 0.191 & 0.0225 & 174.0 & 10.60 & 6.615 & 34.18 & 34.19 & 35.13 \\
$\cdots$ & 40.0 & 0.086 & 0.3769 & 104.3 & 0.103 & 0.073 & 34.13 & 33.60 & 34.24 \\
$\cdots$ & 220.0 & 0.005 & 0.0270 & 51.75 & 3.632 & 2.268 & 34.18 & 32.60 & 33.03 \\
$\cdots$ & 270.0 & 0.667 & 0.0698 & 264.1 & 15.04 & 2.623 & 34.74 & 34.80 & 35.86 \\
$\cdots$ & 300.0 & 0.016 & 0.1635 & 69.61 & 25.11 & 20.32 & 34.07 & 33.62 & 33.54 \\
$\cdots$ & 200.0 & 0.500 & 0.1397 & 91.09 & 3.819 & 0.866 & 34.62 & 33.53 & 34.01 \\
$\cdots$ & 150.0 & 0.527 & 0.0413 & 164.1 & 0.517 & 0.178 & 34.44 & 34.07 & 35.03 \\
$\cdots$ & 360.0 & 0.722 & 0.0068 & 34.69 & 1.100 & 7.013 & 33.17 & 31.78 & 32.33 \\
$\cdots$ & 150.0 & 0.103 & 0.0808 & 63.81 & 2.124 & 2.640 & 33.88 & 32.71 & 33.39 \\
$\cdots$ & 110.0 & 0.161 & 0.4056 & 39.78 & 0.645 & 1.138 & 33.73 & 32.01 & 32.57 \\
$\cdots$ & 190.0 & 0.286 & 0.1179 & 73.94 & 3.485 & 1.673 & 34.30 & 33.03 & 33.65 \\
$\cdots$ & 285.0 & 0.183 & 0.0360 & 69.06 & 2.660 & 2.582 & 33.99 & 32.96 & 33.53 \\
$\cdots$ & 190.0 & 0.394 & 0.1627 & 103.6 & 0.742 & 0.325 & 34.33 & 33.85 & 34.23 \\
$\cdots$ & 75.0 & 0.908 & 0.2547 & 160.0 & 0.259 & 0.034 & 34.85 & 33.98 & 34.99 \\
$\cdots$ & 200.0 & 0.667 & 0.0698 & 264.1 & 13.92 & 2.429 & 34.74 & 34.80 & 35.86 \\
$\cdots$ & 60.0 & 0.546 & 0.2656 & 82.87 & 0.925 & 0.214 & 34.61 & 32.93 & 33.84 \\
 \hline
H15      & 350 & 15.0 & 8.5113 & 83 & 33.0 & 0.300 & 34.9 & 33.4 & 33.9 \\
$\cdots$ & 530 & 24.0 & 36.307 & 108 & 26.0 & 0.175 & 35.1 & 33.4 & 34.3 \\
$\cdots$ & 450 & 37.0 & 3.1622 & 161 & 97.0 & 0.409 & 35.3 & 34.4 & 35.0 \\
$\cdots$ & 1500 & 19.0 & 19.952 & 184 & 39.0 & 0.769 & 35.0 & 33.9 & 35.3 \\
$\cdots$ & 1500 & 8.0 & 213.79 & 115 & 9.0 & 0.376 & 34.6 & 32.8 & 34.4 \\
$\cdots$ & 370 & 10.0 & 13.182 & 52 & 34.0 & 0.308 & 34.7 & 32.8 & 33.1 \\
$\cdots$ & 1500 & 29.0 & 7.7624 & 225 & 74.0 & 0.996 & 35.2 & 34.4 & 35.6 \\
$\cdots$ & 550 & 10.0 & 3.4673 & 72 & 48.0 & 0.942 & 34.7 & 33.4 & 33.6 \\
$\cdots$ & 520 & 11.0 & 3.9810 & 88 & 37.0 & 0.780 & 34.7 & 33.5 & 34.0 \\
$\cdots$ & 360 & 8.0 & 3.2359 & 77 & 30.0 & 0.703 & 34.6 & 33.4 & 33.7 \\
$\cdots$ & 990 & 29.0 & 41.686 & 151 & 30.0 & 0.284 & 35.1 & 33.7 & 34.9 \\
$\cdots$ & 510 & 7.0 & 0.9549 & 94 & 99.0 & 2.003 & 34.5 & 33.8 & 34.1 \\
$\cdots$ & 570 & 9.0 & 2.4547 & 123 & 45.0 & 1.228 & 34.6 & 33.9 & 34.6 \\
$\cdots$ & 370 & 5.0 & 2.0417 & 48 & 3.5 & 1.079 & 34.4 & 33.0 & 32.9 \\
$\cdots$ & 780 & 23.0 & 102.32 & 132 & 15.0 & 0.158 & 35.0 & 33.3 & 34.7 \\
$\cdots$ & 440 & 14.0 & 4.3651 & 94 & 47.0 & 0.559 & 34.8 & 33.6 & 34.1 \\
$\cdots$ & 660 & 27.0 & 51.286 & 88 & 21.0 & 0.178 & 35.1 & 33.2 & 34.7 \\
$\cdots$ & 490 & 6.0 & 6.9183 & 94 & 21.0 & 0.765 & 34.5 & 33.3 & 34.1 \\
$\cdots$ & 700 & 9.0 & 5.6234 & 88 & 35.0 & 0.972 & 34.6 & 33.4 & 34.0 \\
$\cdots$ & 1000 & 36.0 & 60.255 & 132 & 28.0 & 0.216 & 35.2 & 33.5 & 34.7 \\
$\cdots$ & 1260 & 41.0 & 30.902 & 240 & 46.0 & 0.353 & 35.3 & 34.2 & 35.7 \\
$\cdots$ & 150 & 0.83 & 0.2691 & 88 & 4.8 & 3.189 & 33.6 & 33.5 & 34.0 \\
$\cdots$ & 60 & 0.32 & 0.5623 & 30 & 2.3 & 1.418 & 33.2 & 32.2 & 32.1 \\
$\cdots$ & 230 & 5.0 & 0.4073 & 132 & 33.0 & 1.581 & 34.4 & 34.2 & 34.7 \\
\hline
\end{tabular}\\
{
(1) Study, 
(2) outflow velocity \Vout\ (\kms); 
(3) $\rm SFR_{z0.2}$ (\mpy);
(4) $\Sigma_{\rm SFR}$ (\sigmpy) ;
(5) Galaxy maximum rotational velocity (or S$_{0.5}$) (\kms);
(6) ejected mass rate $\dot M_{\rm out}$ (\mpy);
(7) Mass loading factor $\eta$;
(8) Momentum injection rate;
(9) Critical momentum flux;
(10) Critical momentum flux for a shell model.
}
\end{table*}

\begin{table*}
\centering
\caption{Summary of galaxy properties continued}
\label{table:gal_prop2}
\begin{tabular}{lccccccccc}
\hline
Study & V$_{\rm out}$ & $\log{(\rm SFR_{0.2})}$ & $\Sigma_{\rm SFR}$ & $S_{0.5}$ & $\dot M_{\rm out}$ & $\eta$ & $\dot p_\star$ & $\log(\dot p_{\rm crit, c})$ & $\log(\dot p_{\rm crit, s})$ \\
(1) & (2) & (3) & (4) & (5) & (6) & (7) & (8) & (9) & (10) \\
\hline
H15      & 170 & 0.16 & 0.4466 & 55 & 1.0 & 6.324 & 32.9 & 32.7 & 33.2 \\
$\cdots$ & 190 & 6.0 & 2.6302 & 108 & 4.6 & 0.479 & 34.5 & 33.6 & 34.3 \\
$\cdots$ & 630 & 2.8 & 1.1220 & 115 & 22.0 & 3.437 & 34.1 & 33.7 & 34.4 \\
$\cdots$ & 340 & 40.0 & 16.982 & 240 & 12.0 & 0.127 & 35.3 & 34.3 & 35.7 \\
$\cdots$ & 150 & 0.13 & 0.9120 & 68 & 1.0 & 4.242 & 32.8 & 32.6 & 33.5 \\
$\cdots$ & 210 & 2.1 & 1.9498 & 72 & 5.4 & 1.038 & 34.0 & 33.1 & 33.6 \\
$\cdots$ & 230 & 4.8 & 0.2344 & 132 & 30.0 & 2.191 & 34.4 & 34.3 & 34.7 \\
$\cdots$ & 380 & 6.9 & 4.3651 & 151 & 4.6 & 0.688 & 34.5 & 33.9 & 34.9 \\
\hline
B12      & 175.0 & 0.765 & 0.0308 & 92.63 & 0.376 & 0.150 & 34.38 & 33.30 & 35.00 \\
$\cdots$ & 500.0 & 0.147 & 0.0062 & 163.3 & 2.763 & 4.605 & 33.76 & 33.59 & 35.08 \\
$\cdots$ & 300.0 & 1.010 & 0.0690 & 114.5 & 0.394 & 0.087 & 34.63 & 32.89 & 34.90 \\
$\cdots$ & 175.0 & 0.889 & 0.0559 & 82.03 & 1.537 & 0.439 & 34.52 & 33.61 & 34.84 \\
$\cdots$ & 225.0 & 0.936 & 0.1035 & 169.7 & 1.635 & 0.408 & 34.58 & 33.73 & 35.21 \\
\hline
M19      & 105.3 & 0.125 & 0.0408 & 91.54 & 0.005 & 0.008 & 32.81 & 31.07 & 34.02 \\
$\cdots$ & 181.38 & 0.149 & 0.0205 & 173.9 & 0.157 & 0.231 & 32.80 & 32.89 & 35.13 \\
$\cdots$ & 50.98 & 1.204 & 0.4732 & 89.42 & 0.001 & 0.000 & 33.75 & 30.87 & 33.98 \\
$\cdots$ & 286.93 & 0.712 & 0.0965 & 157.1 & 0.955 & 0.369 & 33.53 & 33.36 & 34.96 \\
$\cdots$ & 239.17 & 0.388 & 0.0134 & 74.10 & 0.725 & 3.601 & $\cdots$ & 32.68 & 33.65 \\
$\cdots$ & 308.43 & 0.809 & 0.1078 & 172.6 & 2.995 & 0.853 & 33.60 & 33.88 & 35.12 \\
$\cdots$ & 152.0 & 0.065 & 0.0228 & 117.6 & 0.169 & 0.384 & $\cdots$ & 32.63 & 34.45 \\
$\cdots$ & 155.09 & 0.508 & 0.1015 & 102.9 & 0.253 & 0.140 & 33.42 & 32.63 & 34.22 \\
$\cdots$ & 155.21 & 0.551 & 0.0081 & 86.66 & 0.481 & 3.624 & $\cdots$ & 32.85 & 33.92 \\
$\cdots$ & 317.98 & 2.074 & 3.8574 & 121.3 & 3.085 & 0.040 & 34.01 & 33.50 & 34.51 \\
$\cdots$ & 115.48 & 0.039 & 0.0276 & 144.1 & 0.012 & 0.020 & 32.71 & 31.74 & 34.81 \\
$\cdots$ & 254.21 & 0.535 & 0.0668 & 155.9 & 1.044 & 0.591 & 33.42 & 33.44 & 34.94 \\
$\cdots$ & 146.47 & 1.061 & 0.1706 & 176.7 & 0.061 & 0.010 & 33.70 & 32.55 & 35.16 \\
$\cdots$ & 406.11 & 0.652 & 0.0722 & 175.3 & 2.047 & 0.841 & 33.51 & 33.61 & 35.15 \\
$\cdots$ & 103.14 & 0.327 & 0.0724 & 63.86 & 0.025 & 0.025 & 33.16 & 31.48 & 33.39 \\
$\cdots$ & 233.36 & 0.951 & 0.2375 & 113.0 & 1.476 & 0.330 & 33.65 & 33.36 & 34.38 \\
\hline
P23      & 1204 & 2.046 & 981 & 183.7 & 34 & 0.184 & 35.94 & 31.39 & 35.23 \\
$\cdots$ & 1426 & 1.847 & 281 & 188.1 & 28 & 0.282 & 35.67 & 32.38 & 35.27 \\
$\cdots$ & 2480 & 1.686 & 1519 & 192.6 & 156 & 1.733 & 35.63 & 33.27 & 35.31 \\
$\cdots$ & 1718 & 1.768 & 1074 & 178.0 & 25 & 0.284 & 35.62 & 31.93 & 35.17 \\
$\cdots$ & 2051 & 2.177 & 100 & 251.3 & 16 & 0.070 & 36.03 & 32.78 & 35.77 \\
$\cdots$ & 1842 & 1.815 & 85 & 147.5 & 132 & 1.450 & 35.64 & 33.00 & 34.85 \\
$\cdots$ & 247 & 1.675 & 2 & 232.4 & 14 & 0.225 & 35.47 & 32.95 & 35.64 \\
$\cdots$ & 1138 & 1.933 & 1755 & 169.9 & 49 & 0.324 & 35.86 & 32.13 & 35.09 \\
$\cdots$ & 1728 & 1.982 & 104 & 257.3 & 16 & 0.083 & 35.96 & 31.74 & 35.81 \\
$\cdots$ & 1514 & 1.843 & 652 & 179.4 & 9 & 0.077 & 35.74 & 31.25 & 35.19 \\
$\cdots$ & 1188 & 1.806 & 22 & 155.9 & 28 & 0.333 & 35.60 & 31.19 & 34.94 \\
$\cdots$ & 1421 & 1.765 & 216 & 154.6 & 13 & 0.118 & 35.72 & 32.58 & 34.93 \\
\hline
\end{tabular}\\
{
(1) Study, 
(2) outflow velocity \Vout\ (\kms); 
(3) $\rm SFR_{z0.2}$ (\mpy);
(4) $\Sigma_{\rm SFR}$ (\sigmpy) ;
(5) Galaxy maximum rotational velocity (or S$_{0.5}$) (\kms);
(6) ejected mass rate $\dot M_{\rm out}$ (\mpy);
(7) Mass loading factor $\eta$;
(8) Momentum injection rate;
(9) Critical momentum flux;
(10) Critical momentum flux for a shell model.
}
\end{table*}

\end{document}